\documentclass[12pt,preprint,letterpaper]{aastex}

\newcommand\msa{\hbox{$\rm mag~arcsec^{-2}$}}          
\newcommand\Hseventy{\hbox{$H_0=70\rm ~km~s^{-1}~Mpc^{-1}$}}  
\newcommand\Lambdacos{\hbox{$\Omega_\Lambda=0.7,\Omega_{\rm m}=0.3$}}
\newcommand\lstar{\hbox{$\rm L^*$}}                          
\newcommand\Msun{\hbox{$\rm M_\odot$}} 
 
\newcommand\etal{{\it et al. }}         
\newcommand\cola {\null}
\newcommand\colb {&}
\newcommand\colc {&}
\newcommand\cold {&}
\newcommand\cole {&}
\newcommand\colf {&}
\newcommand\colg {&}
\newcommand\colh {&}
\newcommand\coli {&}
\newcommand\colj {&}
\newcommand\colk {&}

\newcommand\eol{\\}

\slugcomment{}

\shorttitle{}
\shortauthors{}

\begin{document}

\title{Host Galaxies of z=4 Quasars
\footnote{Based on data obtained with the 6.5 meter Baade Telescope
of the Magellan Telescopes, located at the Las Campanas Observatory, Chile.}
\footnote{Based on observations obtained at the Gemini Observatory,
which is operated by the Association of Universities for Research in
Astronomy, Inc., under a cooperative agreement with the NSF on behalf
of the Gemini partnership: the National Science Foundation (United
States), the Particle Physics and Astronomy Research Council (United
Kingdom), the National Research Council (Canada), CONICYT (Chile), the
Australian Research Council (Australia), CNPq (Brazil) and CONICET
(Argentina).}
\footnote{Based in part on data taken at the W. M. Keck Observatory, which is
operated as a scientific 
partnership among the California Institute of Technology, the
University of California, and 
NASA, and was made possible by the generous financial support of the
W. M. Keck Foundation.}
}

\author{K. K. McLeod}
\affil{Whitin Observatory, Wellesley College, Wellesley, MA 02481}
\email{kmcleod@wellesley.edu}
\and
\author{Jill Bechtold}
\affil{Steward Observatory, University of Arizona, Tucson, AZ 85721}
\email{jbechtold@as.arizona.edu}

\begin{abstract}

We have undertaken a project to investigate the host galaxies and
environments of a sample of quasars at $z\sim4$.  In this paper, we
describe deep near-infrared imaging of 34 targets using
the Magellan I and Gemini North telescopes.  We discuss in detail
special challenges of distortion and nonlinearity that must be
addressed when performing PSF subtraction with data from these
telescopes and their IR cameras, especially in very good seeing.  We
derive black hole masses from emission-line spectroscopy, and we calculate accretion 
rates from our $K_s$-band photometry, which directly samples the
rest-frame $B$ for these objects. We introduce a new isophotal
diameter technique for estimating host galaxy luminosities. We report
the detection of four host galaxies on our deepest, sharpest images, 
and present upper limits for the
others. We find that if host galaxies passively evolve such 
that they brighten by 2 magnitudes or more in the rest-frame $B$ band 
between the present and z=4, then high-z hosts are less
massive at a given black hole mass than are their low-z counterparts.
We argue that the most massive hosts plateau at $\lesssim10\lstar$.
We estimate the importance of selection effects on this survey and the 
subsequent limitations of our conclusions.
These results are in broad agreement with recent semi-analytical models 
for the formation of luminous quasars and their host spheroids by mergers of 
gas-rich galaxies, with significant dissipation, and self-regulation of
black hole growth and star-formation by the burst of merger-induced 
quasar activity.  

\end{abstract}


\section{Introduction}

In the past decade, we have begun to understand the important role that
black holes play in galaxy evolution.  Observations suggest that
supermassive nuclear black holes are likely present in nearly all normal
galaxies, and that black hole mass is correlated with
host galaxy bulge mass \citep{kor95, mag98} and stellar velocity
dispersion \citep{fer00, geb00, tremaine2002, mar03, har04}.
Low-redshift quasars too are consistent with these results: at low redshift, 
the most luminous quasars reside in massive, early-type host galaxies,
and fit the black-hole-mass-spheroid relation for Eddington fractions
of about $\sim40\%$ \citep{mr95, mrs99, mcl99, mm01, mclure2001,
  flo04, kiuchi2009, silverman2009}.
These results have implications for the evolution of luminous,
high-redshift quasars.  If a galaxy already had a supermassive black
hole early on, then according to the local black-hole/bulge relation,
it must by today be one of today's most massive galaxies. 

In the context of the $\Lambda$CDM 
framework for hierarchical structure growth, 
specific predictions can be made for the evolution of 
quasar hosts and their environments
through cosmic time \citep{mo02, kh02}. 
If dissipationless gravitational collapse of cold dark matter 
were the only process at
work, then one would expect the ratio of black hole mass 
to stellar spheroid mass, $M_{BH}/M_*$, to be roughly 
constant as spheroids merge and their nuclei coalesce.  
However, luminous quasars like the ones
in current samples of quasars at $z\ge 4$ are likely the product of 
major mergers of gas-rich disk
galaxies of comparable mass \citep{hopkins2005, croton2006, dimatteo2008, hopkins2008, somerville2008}.  The central black holes merge, and 
merger-induced gas accretion results in a burst of 
quasar activity.  Quasar radiative energy and winds eventually halt 
further mass accretion and clear out the cold gas, halting star-formation.
The merger disrupts 
the original gaseous disks, and the result is a low angular-momentum 
spheroid of stars that subsequently evolves passively.
Semi-analytical models for these processes, along with numerical simulations 
of the collapse of cold-dark-matter halos, are successful in reproducing 
many observations of galaxies, galaxy clusters and quasars.  
In these models, high-z quasars are expected to have less luminous 
hosts than their low-z
counterparts \citep{kh00}, with the ratio of black hole mass to
stellar spheroid mass, $M_{BH}/M_*$, decreasing with redshift
\citep{croton2006, somerville2008}.

By necessity, models for quasar host evolution rely on semi-empirical 
prescriptions for key physical processes, because the resolution of
numerical simulations cannot follow all of the crucial physics 
from the spatial scales of galaxy clusters down to galactic then to atomic scales.  
Direct observations of high redshift quasar hosts such as the study described here 
provide one interesting
empirical check on the overall validity of the theoretical picture of
hierarchical  
galaxy and black hole formation and evolution.

Detecting the host galaxy ``fuzz'' is technically challenging
at high redshift however because it appears 
small and faint compared to scattered light from the nucleus in the
wings of the point spread function (PSF). Ideally, we would
study the fuzz in the rest-frame near-IR, which would both
highlight the mass-tracing stellar populations of the hosts and
provide the best possible galaxy-to-nuclear light contrast \citep{mr95}.  
For high-z objects, that would mean observing in the mid-IR, but there are
not yet telescopes with the necessary combination of sensitivity and angular
resolution to make such observations feasible.  Most high-z host studies so 
far have therefore used near-IR imaging. 

At $z\sim2-3$, the first near-IR imaging of handfuls of objects
using 4m-class telescopes produced detections only in the case of
radio-loud (RL) quasars \citep{lehnert92,lowenthal95,aretxaga98a,carballo98}.  
Fuzz was subsequently seen around a few radio-quiet (RQ) quasars using
adaptive optics (AO) on 4m telescopes
\citep{aretxaga98b,hutchings99,kuhlbrodt2005}, and AO on 
the Gemini North 8m yielded only one of 9 hosts at $z\sim2$
\citep{croom2004}.
The well-characterized and stable PSF of NICMOS allowed successful
detections of larger samples of RQ hosts in this 
redshift range \citep{kukula2001,ridgway2001,peng2006}, which led to the first
tantalizing comparisons with hierarchical models.  
\citet{peng2006} suggest that at
$z=2$,  $M_{BH}/M_*$ was several times larger than it is today, 
and that hosts were not yet fully formed, though those results have
been the subject of some debate. For example \citet{falomo2008} have recently
used AO imaging on the VLT to detect the hosts of three luminous
quasars at $2<z<3$, which they use to help argue for passively
evolving $\sim5\lstar$ elliptical hosts.   Moreover, Malmquist bias in 
high-z samples would skew detections towards bright quasars in small hosts 
\citep{lauer2007,treu2007,woo2008}; we discuss this effect further in
\S\ref{sec-malmquist}.

To provide more leverage for probing the hierarchical models, we
would like to measure host properties at $z\sim4$.  
So far, host detections have been claimed for just a few objects in
this range, all with unknown radio type.  \citet{peng2006} used
NICMOS to measure host magnitudes and sizes for two gravitationally
lensed quasars at $z=4.1$ and 4.5. Hutchings has used Gemini to
observe seven targets with $z\sim5$
\citep{hutchings2003,hutchings2005}; unfortunately, the 
results are
hard to interpret in light of the distortion and nonlinearity effects
that we have found to be significant for the same instrument (see
below). 

With the improvement in sensitivity provided by 6-8m class telescopes,
we have begun to expand the sample of quasars imaged at high z.
Our long-term program includes new multi-wavelength imaging to search
for hosts and to characterize environments; new and archival visible
spectroscopy to be used for virial mass estimates of the central
engines from emission line properties; and modeling of the accretion
disks and hot coronae using data from IR to X-ray, including new X-ray
data from ongoing Chandra and XMM observing programs.

In this paper, we describe a sample of 34 $z\sim4$ quasars that we have imaged 
in the near-IR.  We present the observations and report on the search
for hosts.  The environments will be described in a subsequent paper (Bechtold \& McLeod 2010, in preparation).
We adopt the cosmology \Lambdacos, \Hseventy\ throughout. 
All magnitudes are reported in the Vega system, unless otherwise noted.

\section{The Sample}

\begin{figure}
\epsscale{0.7}
\plotone{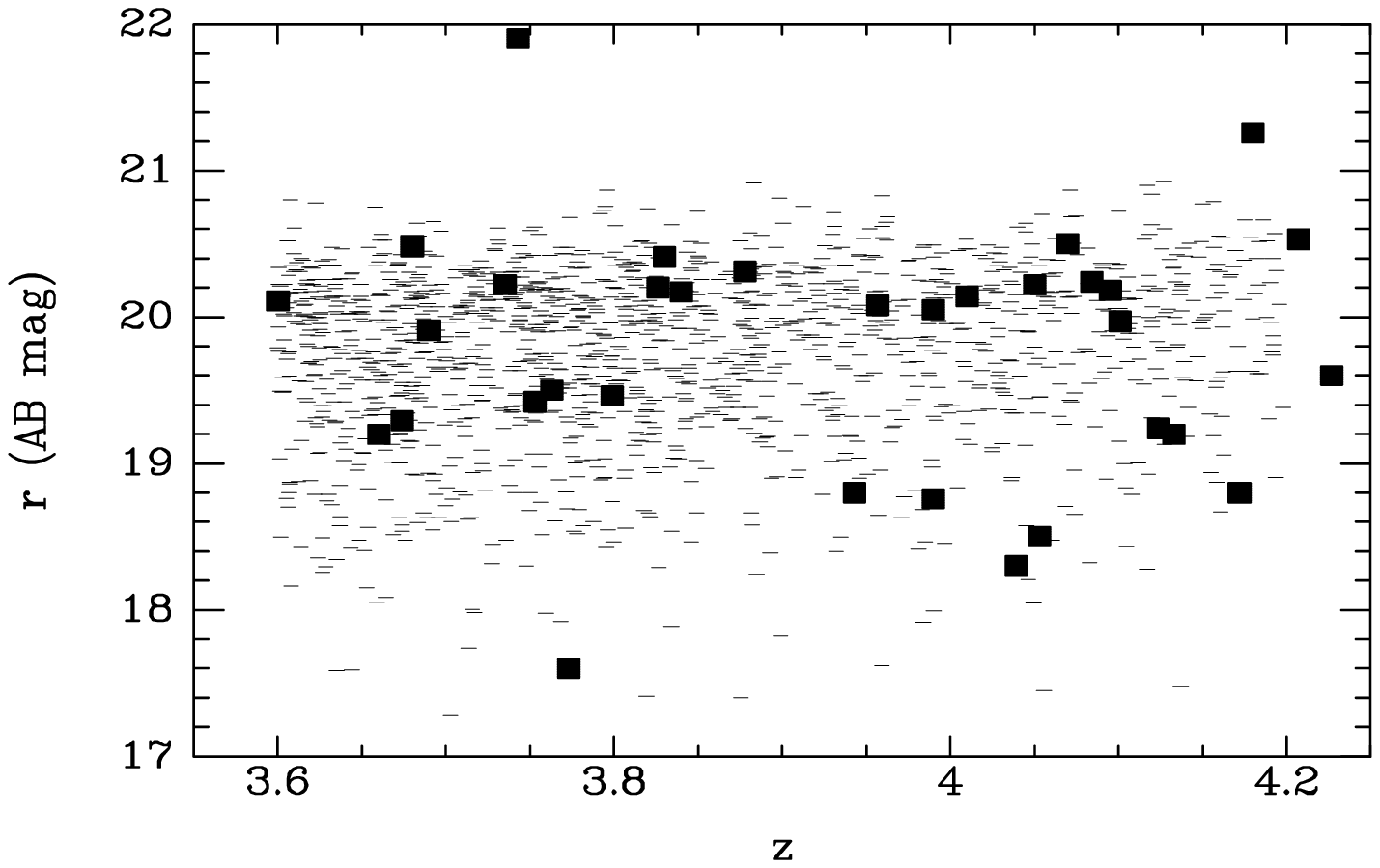}
\caption{Magnitude and redshift distribution for the 34 quasars
observed here (squares) and quasars from the Sloan Quasar Catalog \citep{schneider2007}.    
\label{fig-sample}}
\end{figure}

We have observed a sample of quasars selected to have redshifts in the
range $3.6\lesssim z \lesssim4.2$.  The sample is listed in Table
\ref{tab-obs} with names as given in the NASA/IPAC Extragalactic
Database (NED).  The redshift range was chosen so that the 4000\AA\
break falls between observed $H$- and $K$-bands, so that broad-band colors give 
maximum leverage for estimating photometric redshifts and stellar populations. 
Out of the $\sim300$ quasars known in this interval when we began the
project,  we observed a randomly-chosen sub-sample of 34 
objects, yielding a median $<z>=3.9$ and spanning a range of 
magnitude as shown in Fig. \ref{fig-sample}. We plot our sample
against the $\sim1600$ quasars at these redshifts listed in the most
recent Sloan Digital Sky Survey (SDSS\footnote{Funding for the Sloan
Digital Sky Survey (SDSS) has been provided by the Alfred P. Sloan
Foundation, the Participating Institutions, the National Aeronautics
and Space Administration, the National Science Foundation, the
U.S. Department of Energy, the Japanese Monbukagakusho, and the Max
Planck Society. The SDSS Web site is http://www.sdss.org/.}) Quasar
Catalog \citep{schneider2007}.  

We are observing first in the $K$-band, which samples the rest frame
$B$ for these objects.   The median observed total nuclear $K_s=17.2$
for the quasars in our sample 
corresponds to $M_B=-26.9$ (Vega magnitudes), similar to the local
luminous quasar 3C273.

\subsection{Radio Loudness}\label{sec-radio}

To characterize the radio properties of our sample, we adopt the definition of radio loudness given in \cite{ivezic2002},
who analyzed the radio properties of SDSS qusars.  They define the apparent AB
magnitude \citep{oke1983}  
at 1.4 GHz as

$$t \equiv -2.5 log_{10} \left({{F_{int}\over{3631 Jy}}}\right)$$

\noindent
where $F_{int}$ is the integrated 20cm radio flux measured from a
two-dimensional Gaussian fit to the radio source.  The
radio-to-optical flux density is then defined as 

$$R_i \equiv 0.4 (i^{AB} - t)$$

\noindent
where $i^{AB}$ is the AB magnitude at Sloan $i$ in the continuum.  
With these definitions \cite{ivezic2002} find that radio-loud quasars 
have $R_i \cong 1 - 4$ and radio-quiet quasars have $R_i < 1$.

The radio properties for the quasars in our sample are given in Table \ref{tab-radio}.  
We compiled the optical magnitudes from a number of sources.  
When available, we adopted the values for $i^{AB}$ published by members of 
the Sloan consortium;  references are listed in 
Table \ref{tab-radio}.  For most other objects, we used the
$i^{AB}$-band photometry  
given on the SDSS web site (DR6).  For two objects we measured photometry from 
archival HST images.  For these and a few other cases, the only available 
photometry was from the literature in other filters, which we transformed 
to $i^{AB}$ using the zero-points given in the NICMOS web site unit converter, 
and assuming a quasar spectral energy distribution in the form of a power law 

$F_{\nu} \propto \nu ^{\alpha}$ where the spectral index
$\alpha = -0.44 $ for $1216\AA \le \lambda_{rest} < 5000 \AA$

\noindent which is the average derived from the SDSS quasars 
by \cite{vandenberk2001} (see also \citet{pentericci2003}).

We corrected the optical magnitudes for Galactic reddening, using the 
$E(B-V)$ given by \cite{schlegel1998} as tabulated in NED.  The 
transformations $A_r = E(B-V) / 2.751$ and $A_i = E(B-V) / 2.086$ were
adopted \citep{schneider2003}. 

For most of the sample quasars, the most sensitive radio data comes from the
Faint Images of the Radio Sky at Twenty-cm Survey \citep[FIRST]{first}, which we accessed through NED.  If the quasar did not fall in one of the FIRST fields, we used the data from NRAO VLA Sky Survey (NVSS; \citep{nvss}).  In a few cases 
the most sensitive radio data had been reported in targeted searches in 
the literature.  For objects detected in FIRST, we adopted the FIRST catalog 
integrated flux.  For others, we derived a 2$\sigma$ upper limit from the 
root-mean-square fluxes in the maps downloaded from NED.  

Of the 34 sample quasars, 16 are radio-quiet, 5 are radio-loud, 4 have no 
radio data, and 9 have radio data which are not deep enough to know whether
or not the quasar is radio-loud or radio-quiet.  

Going into our survey, we wanted to test whether radio-loud quasars have
different host galaxy properties than the radio-quiet majority.  We therefore 
tended to give priority to observing quasars which we knew to be radio-loud, since 
they are rare and we knew that it would be difficult to get a statistically 
large sample of them.  In the end, at least 5 quasars in
the sample of 34 are radio-loud, compared to approximately 1 expected to
be radio-loud, had we observed a sample representative of the radio 
properties of the bright quasar population at $z\sim4$ as a whole
\citep{jiang2007}.   

\subsection{Black Hole Mass Estimate}

We estimated the black hole mass $\rm M_{BH}$ for the quasars in our
study from emission-line spectroscopy.  As described below, we measured the full-width-half-maximum (FWHM)
value of the broad CIV emission line and the quasar UV continuum luminosity from new and existing spectra.  We then used these to compute black hole mass according to the relation 

$$log_{10}\left({\rm M_{BH}\over M_\sun}\right) =
log_{10}\left[\left({ {\rm FWHM_{CIV}} \over{1000km~s^{-1} }}\right)^2 \left({\lambda L_{\lambda1350}
\over{10^{44}erg~s^{-1}}}\right)^{0.53}\right]+6.66$$

\noindent from \cite{vestergaard2006}, who find that the UV continuum luminosity $L_{\lambda1450}$ can be freely substituted for $L_{\lambda 1350}$.  From this we derived

$$log_{10}\left({\rm M_{BH}\over M_\sun}\right) =
2 log_{10}\left({{\rm FWHM_{CIV}}\over{1000km~s^{-1}}}\right)
+ 0.53 \left[11.37 + {2*\rm {DM}\over{5}} -{{\rm {AB}_{1450}+48.60}\over{2.5}} -log_{10}(1+z)\right] 
+ 6.66$$

\noindent Here, ${\rm AB_{1450}}\equiv-2.5log_{10}f_\nu-48.6$, where
$f_\nu$ is the (reddening-corrected) continuum flux in
$erg~s^{-1}cm^{-2}Hz^{-1}$ measured at an observed wavelength of
$\lambda =1450\AA(1+z)$ as in \cite{fan01}.  $DM$ is the (luminosity)
distance modulus.   

The $\rm AB_{1450}$ magnitudes were compiled from the literature or
measured by us as shown in Table \ref{tab-blackholes}.  Where
available, we adopt the reddening-corrected $\rm AB_{1450}$ tabulated
by members of the Sloan consortium, who give values based on
spectrophotometry of the 
quasars at a rest frame wavelength of 1450\AA.  For 10 of the objects, we
measured the continuum fluxes ourselves from spectra obtained from the
SDSS Skyserver.  For the objects for which no flux-calibrated spectra
were available, we used the $i^{AB}$ magnitudes from Table \ref{tab-radio}
and transformed to $\rm AB_{1450}$ assuming $\alpha=-0.44$ as described
in \S \ref{sec-radio}.  For objects with large CIV equivalent widths
that contaminate the broadband measurements, the $\rm AB_{1450}$
derived from photometry will be systematically bright.  A comparison
of the spectroscopically-derived $\rm AB_{1450}$ to the
photometrically-derived one for the SDSS objects shows the former to
be fainter on average by $0.4\pm0.3$ mag. 

\begin{figure}
\epsscale{1}
\plotone{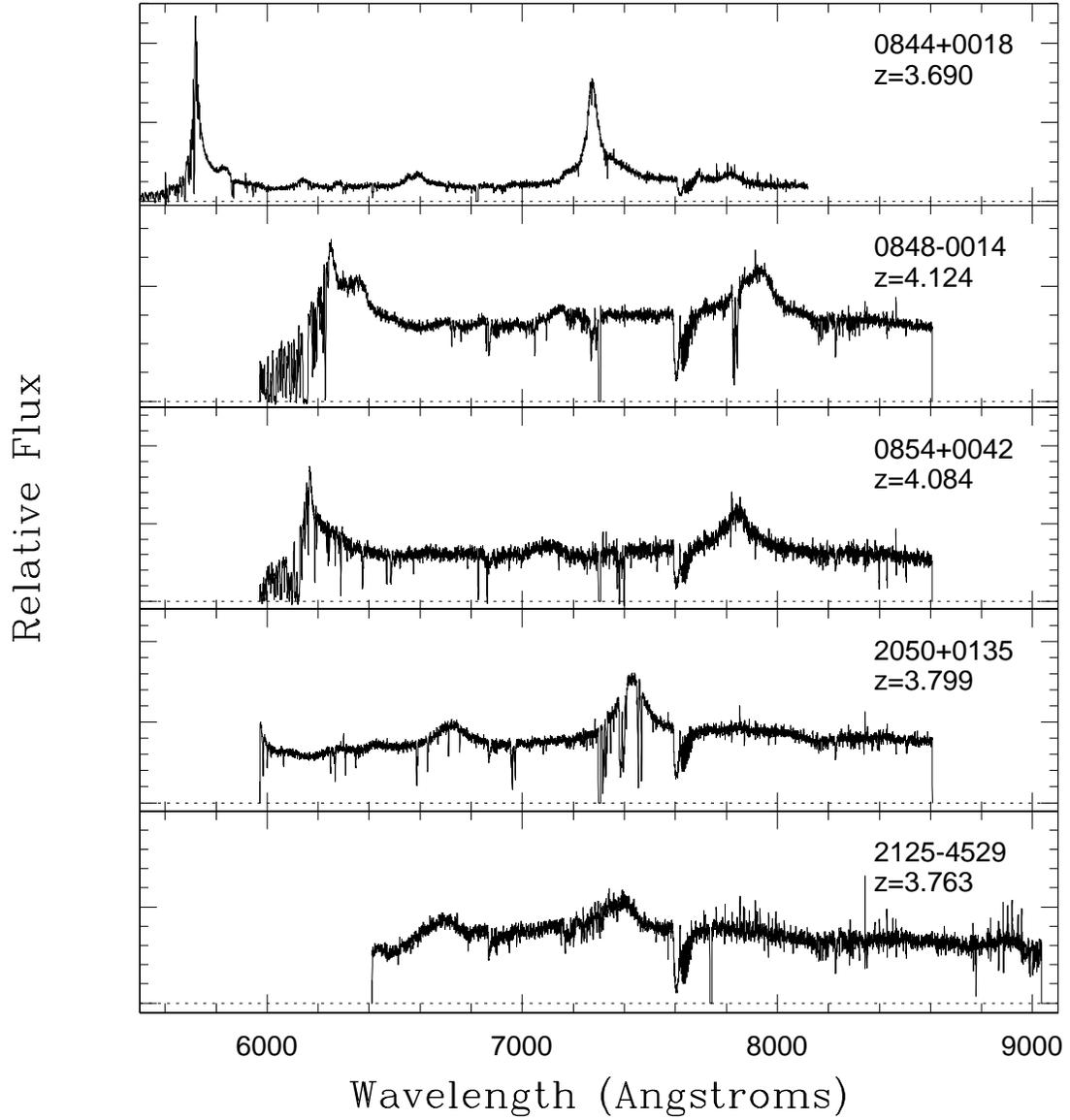}
\caption{New Keck DEIMOS spectra of five quasars in the region around CIV in
the rest-frame used for estimating black hole masses. The spectrum for
[VH]2125-4529 also provided an improved redshift.
\label{fig-spec}}
\end{figure}

For most objects, we measured the FWHM of the C~IV emission lines
given in Table \ref{tab-blackholes} using spectra from the SDSS
Skyserver or electronic versions of published spectra from several
authors who kindly made them available.  In a few cases, we digitized
published spectra using 
$``$Plot Digitizer" software.  
We also carried out new
long-slit optical spectroscopy 
for five targets in the sample, including one
for which no other spectroscopy is published, [VH95]2125-4529.   For 
the new observations, we used the
DEIMOS spectrograph on the 
Keck-II telescope  \citep{davis2003} on the nights of 2008 Oct 24 and
2004 Oct 12 and 13.  The 
objects were observed through a 0.7 arcsec wide slit with the 1200 l/mm
first order grating,  
resulting in a dispersion of 0.33 $\AA$/pixel.  A GG495 filter was used to
block second order light.
Exposures were 600-900 seconds, mostly
through clouds or at twilight.  We reduced the data using the DEEP2
project IDL reduction pipeline,
which flatfielded, sky-subtracted, wavelength-calibrated and extracted
the spectra as 
described in the DEEP2 webpage,
http://astro.berkeley.edu/$\sim$cooper/deep/spec2d/.   Spectra
are shown in Figure \ref{fig-spec}.  

To derive the CIV line width, we subtracted a local continuum fit,
derived by fitting a linear curve through the spectra in rest
wavelengths 1425\AA~ to 1500\AA~ and 1760 \AA~ to 1860 \AA.    
We replaced absorption features with an interpolated continuum
estimate, and then fit a gaussian to the C IV emission line.

The quality of the C~IV line profiles for the quasars in our sample
ranged from very high-signal-to-noise examples with easy-to-define
continua, to barely detected lines in discovery-quality spectra.
Moreover, the redshifts of our targets shift 
the C~IV line to wavelengths with strong telluric absorption and 
night-sky emission features that are difficult to calibrate out
completely.  Some lines probably are suppressed by undetected
absorption features intrinsic to the quasar.  As many authors have
noted, quasar emission lines are non-Gaussian, in the sense that they
have ``pointy'' peaks.  Some C~IV lines in our sample were also
significantly asymmetric.   

For these reasons, we measured the FWHM values by hand, using
IRAF's\footnote{IRAF (Image Reduction and Analysis 
Facility) is distributed by the National Optical Astronomy
Observatories, which are operated by the Association of Universities
for Research in Astronomy, Inc., under contract with the National
Science Foundation.}   {\it splot}.  In cases where  part of the line
profile was very noisy, we 
measured the half-width of the better side of the profile and doubled
it.  In a few of the spectra with very good signal-to-noise, 
there is clearly a narrow (2000-3000 km s$^{-1}$ FWHM) component, and
broader wings (10,000-15,000 km  s$^{-1}$ FWHM).  For several objects, we
estimated two values of the line width, one for each component; both
are listed in Table \ref{tab-blackholes}. For [VCV96]Q2133-4625, the 
C IV profile is so noisy, possibly because of an absorption trough,
that it was  
impossible to derive a FWHM from the published spectrum.

Our resulting black hole mass estimates are given in Table
\ref{tab-blackholes}. 
The systematic uncertainties in these estimates for black hole mass are well-known \citep{wandel1999, 
collin2002, dh2004, kaspi2005, vestergaard2006, netzer2007, kb2007, mcgill2008}.  The primary assumption is  
that the CIV emitting gas is in virial equilibrium
with the central black hole mass, and is located at a radius that
scales with luminosity.  For the luminous quasars in our sample, this
means extrapolating from the relations tested in
emission-line regions studied with reverberation mapping locally 
\citep{peterson2004}.  Further, the
bolometric luminosity of each quasar is assumed to be a constant
multiple of the $\lambda$1450 continuum luminosity, which certainly is
not the case (e.g. \citet{kelly08}).    

\begin{figure}
\epsscale{0.7}
\plotone{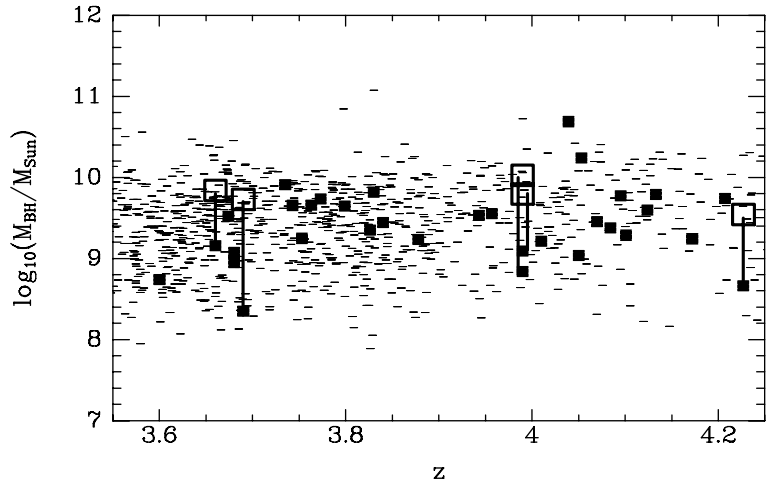}
\caption{Black hole masses as a function of z for the 34 quasars
observed here (filled squares) shown on top of those determined for $\approx
1600$ SDSS quasars in this redshift range by \cite{shen2008}.   For
several objects the CIV line profile shows a distinct broad component
whose FWHM yields black hole masses (open squares) an order of magnitude
larger than the overall FWHM would indicate.  
\label{fig-masses}}
\end{figure}

In Figure \ref{fig-masses} we plot the black hole masses of the
quasars in our sample along with those for the $\approx1600$ SDSS
quasars in this redshift range recently tabulated by \cite{shen2008}.
For the 6 objects in common, our black hole  
mass estimates generally agree within $\approx0.2dex$.
 
\subsection{Accretion Rates}

We combined the black hole masses with the K-band observations
described below to calculate the quasar mass accretion rates.  Because the
observed K-band samples the rest-frame 
B-band, the K-band magnitude allows us to compute a B-band luminosity
{\it independently of spectral shape}.  This avoids the errors
that result when one must extrapolate from optical photometry to the
rest-frame B assuming a spectral index $\alpha$.  We
apply a B-band bolometric correction factor of 10.7 \citep{elvis94}
and we compare the resulting bolometric luminosity to the Eddington
luminosity computed from the black hole mass via
$L_{Edd}=3.3\times10^4({\rm M_{BH}}/M_\sun)~L_\sun$.  We have assumed
that all of the rest-frame B-band light can be attributed to the
nucleus, which is a reasonable estimation for such luminous objects.  The
resulting accretion rates as fractions of Eddington, $L_{bol}/L_{Edd}$, are tabulated in Table \ref{tab-blackholes}. 

The median $L_{bol}/L_{Edd}$ for the sample is $0.47\pm1.6~(1\sigma)$, and the
minimum value is 0.1. These rates are good matches to those inferred
from studies of host galaxies locally;  \cite{mm01} found that the
most luminous local quasars radiate at $\gtrsim0.1L_{Edd}$, and
\cite{flo04} deduce a median rate $0.47$ for the most luminous local
quasars.  

The calculation of accretion rates yields a
handful of quasars with super-Eddington rates. 
Of these, 
BR2212-1626 is gravitationally lensed \citep{warren2001} and so the continuum
luminosity, which we have not corrected for gravitational magnification,
is overestimated.  Since both $M_{BH}$ and $L/L_{Edd} \propto L^{0.5}$,
both quantities are also overestimated.  For BRJ0529-3553, we have only a 
discovery quality spectrum, and the C IV line width is very uncertain.
For 5 others which have $L/L_{Edd} >>1$, the C IV profile has good enough 
signal-to-noise to detect a distinct narrow and broad component.  If the
FWHM of the broad component is used, very large black hole masses, and sub-Eddington 
accretion rates are implied.  Detailed modeling of the quasar spectra energy 
distribution and higher quality spectra of all targets would improve the estimates of 
black hole mass and accretion rate.  We do not list the statistical errors for
$M_{BH}$ and accretion rate in Table \ref{tab-blackholes} because these numbers are dominated by
systematic uncertainties and the simplifying assumptions described above.

Excluding the $L/L_{Edd}>1$ objects, the median rate becomes
$0.41\pm0.3$, consistent with the distribution plotted by
\cite{shen2008} for $z>3$ SDSS quasars. 

As a second way to estimate black hole masses for our sample quasars,
we assume that all of the quasars are radiating at $0.4L_{Edd}$, with
bolometric luminosities determined from $M_B$ as above.  The derived
values are then the minimum plausible black hole mass that the nuclei
could have to be emitting at the luminosity observed.  These values
are listed in Table \ref{tab-blackholes}.

\section{Near-IR Imaging Observations}

We have obtained deep, near-IR images of 34 quasars over the period
2002 September - 2005 January at the Magellan I 6.5m and Gemini North
8m telescopes.  We have observed each field in $K_s$, with 5 also
observed in $H$ or $H_c$.  Most of the objects (26) were observed with
Magellan's PANIC \citep{mar04}, a $1024\times1024$ HgCdTe array with a
pixel scale of 0\farcs125 and a field-of-view (FOV) of 128\arcsec.  Before PANIC was
installed, we imaged a few objects (6) with the old ClassicCam
\citep{per92}, a $256\times256$ HgCdTe array camera yielding a 
FOV of only $\sim30\arcsec$ per exposure.  The rest of our
targets (7) were observed on Gemini with NIRI \citep{hod03}, a
$1024\times1024$ InSb array operated at f/6, yielding a pixel scale of
0\farcs116 and a FOV of 119\arcsec\ per exposure.  The observations
are summarized in Table \ref{tab-obs}.

With all three instruments, we observed using a 9- or 25- point dither
pattern of short exposures ($10-30$ sec each, repeated $1-3$ times per
dither position).  The times were chosen to keep the quasar images in the
nominal linear range, with repeats limited to ensure fair sampling of sky
variation.  The dithers were typically repeated for half a night,
yielding up to a thousand frames per field and average total on-source
times of 3 hours for PANIC and NIRI.  With NIRI and PANIC,
the FOV was big enough for us to use stars on the quasar frames to measure the PSF. 
Due to ClassicCam's smaller FOV, we had to alternate quasar dithers with dithers on a nearby star to sample the PSF.  The average on-source time for ClassicCam was thus shorter, 
only about 2 hours, resulting in shallower images.  As we describe in the sections below, the ClassicCam images turned out to be of very limited use for the host searches.  However, we include them in this paper both because they remain somewhat useful for investigations of the near environments of the quasars, and to illustrate the difficulties of using out-of-field stars for PSF subtraction.

We were fortunate to have excellent seeing for many of the
observations, with the final, combined $K_s$ quasar images from PANIC and
NIRI having full-width half-max (FWHM) ranging from
$0\farcs32-0\farcs66$ with median value $<FWHM>=0\farcs43$.
ClassicCam images were worse. 
Photometric calibration was done for the PANIC and NIRI images using
the 2MASS stars (usually $1-3$) found on each combined quasar frame.
We also cross-checked these values against observations of Persson
faint IR standards 
\citep{per98}, and estimate that the photometric calibration is
accurate to $\sim0.1\rm mag$.  For the ClassicCam images, whose FOV are 
too small to contain 2MASS stars, we used only the Persson \etal\
standards.

\section{Data Reduction}

For all three instruments, we reduced the data using standard
techniques in IRAF with its add-on packages {\it gemini/niri} and {\it
panic}, the latter kindly provided by Paul Martini.  The data from
each instrument were handled somewhat differently, with flats made
from twilight exposures, flat lamp exposures, and object frames for
PANIC, NIRI, and ClassicCam respectively.  Sky frames were generally
made by median-filtering $9-10$ dither positions after masking out
sources.  For the NIRI images, persistence proved to be a significant
problem, which we solved to our satisfaction by including in each
frame's bad pixel mask the object masks from the 2 previous frames.   

The PANIC and NIRI detectors are known to be nonlinear by
{$\approx1\%$} at 15,000ADU and 11,000ADU respectively, and we 
kept our quasar+sky counts well below these limits.  Still, for the
tight tolerances in this project we needed to take some care with the
linearity correction.  For PANIC we used flat lamp exposures of
varying lengths to determine our own second-order correction {which
differed from the nominal pipeline correction by 0.5\% at 15,000ADU}.
The NIRI pipeline does not offer any nonlinearity correction and we
lacked the data to determine our own.  We discuss the residual effects
of nonlinearity in \S\ref{sec-psf}. 

In the course of our analysis we detected a geometric distortion in
the PANIC images.  The distortion was visible as a radial stretch in
contour plots of stars taken around an image.  Paul Martini gave 
us a second-order geometric distortion correction derived from the
PANIC optical prescription, which we then implemented in the pipeline.
We note that the NIRI pipeline does not offer a distortion correction,
though distortion proved to be an issue there as well.  We discuss the
implications further in \S\ref{sec-psf}. 

\begin{figure}
\epsscale{1}
\plotone{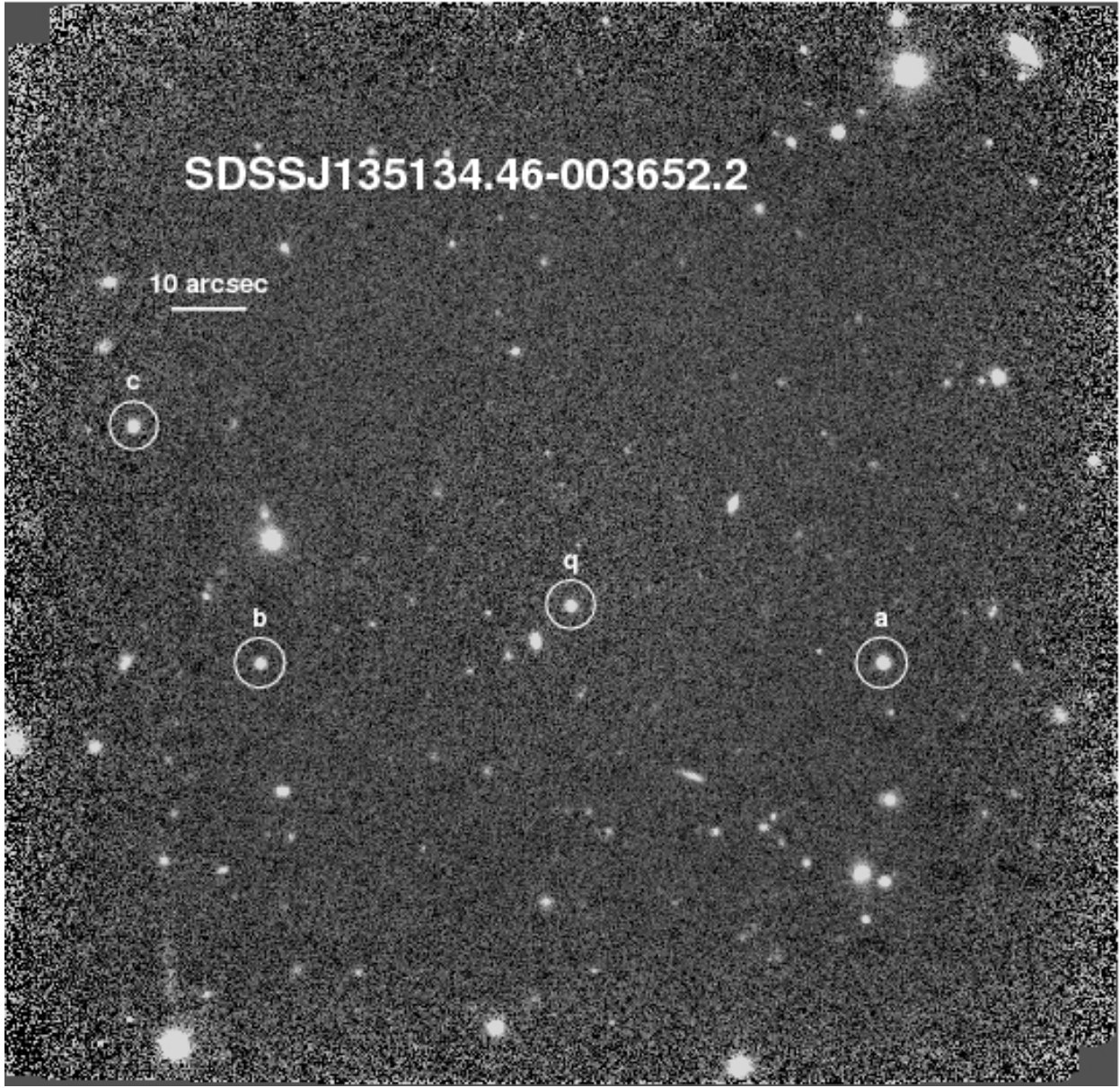}
\caption{Sample PANIC image showing a $z=4.0$ quasar ``q" and PSF stars ``a" and ``c."  The surface brightness limit in the center is $K=22\msa$, which is slightly better than average, but at $0\farcs5$ the seeing is slightly worse than average.  At this redshift, the field
shown is $\sim1\rm Mpc$ across.\label{fig-bigimage}}
\end{figure}

The hundreds of reduced frames for each quasar were magnified by a
factor of two, aligned on the quasar centroid, and combined after
rejecting a handful of frames deemed bad because of bias level
jumps in one quadrant or poor flattening.  
The deepest NIRI $K_s$
images reach a surface brightness limit of $K_s=22.9 \msa$ (measured
as a $1\sigma$ pixel-to-pixel variation).  The median value is 21.7
\msa, typical for the PANIC images, while the ClassicCam images are
more shallow. 
A typical PANIC image is shown in Fig. \ref{fig-bigimage},
where the FOV corresponds to $\sim1.3\rm \ Mpc$ at $z=4$, well-suited
for studies of quasar environments (see Bechtold \& McLeod 2009).

\section{PSF Characterization}\label{sec-psf}

Any search for host galaxies is only as good as the characterization
and removal of the nuclear point source.   We followed the traditional
practice of selecting ``PSF stars'' from each image, and using them in
model fits to the quasar images. 
However, in the course of our analysis we discovered some subtle
effects of residual distortion and nonlinearity.  Because we have not
seen these issues addressed in other high-z host searches, including
ones also done with NIRI, we discuss them in some detail here.  For a
recent look at the PSF perils that host galaxy studies might encounter
even with HST, see \cite{kim2008}. 

\subsection{Geometric distortion, or What to Do When The Seeing is Too Good}

Beginning with PANIC, our experiments with multiple PSF stars showed
that poorer fits tend to result when the PSF star is farther from the
quasar. 
Even though we had performed a distortion correction on the PANIC
images, we found that {\it a small residual geometric distortion
compromised the fits}.  The distortion we detected would be
insignificant (and indeed not noticeable) for most projects, with
camera optics generally designed to create instrumental PSFs small
compared to the seeing size.  In our case, however, the excellent
seeing and tight tolerances required for high-z host detection made
the distortion apparent.

\begin{figure}
\plotone{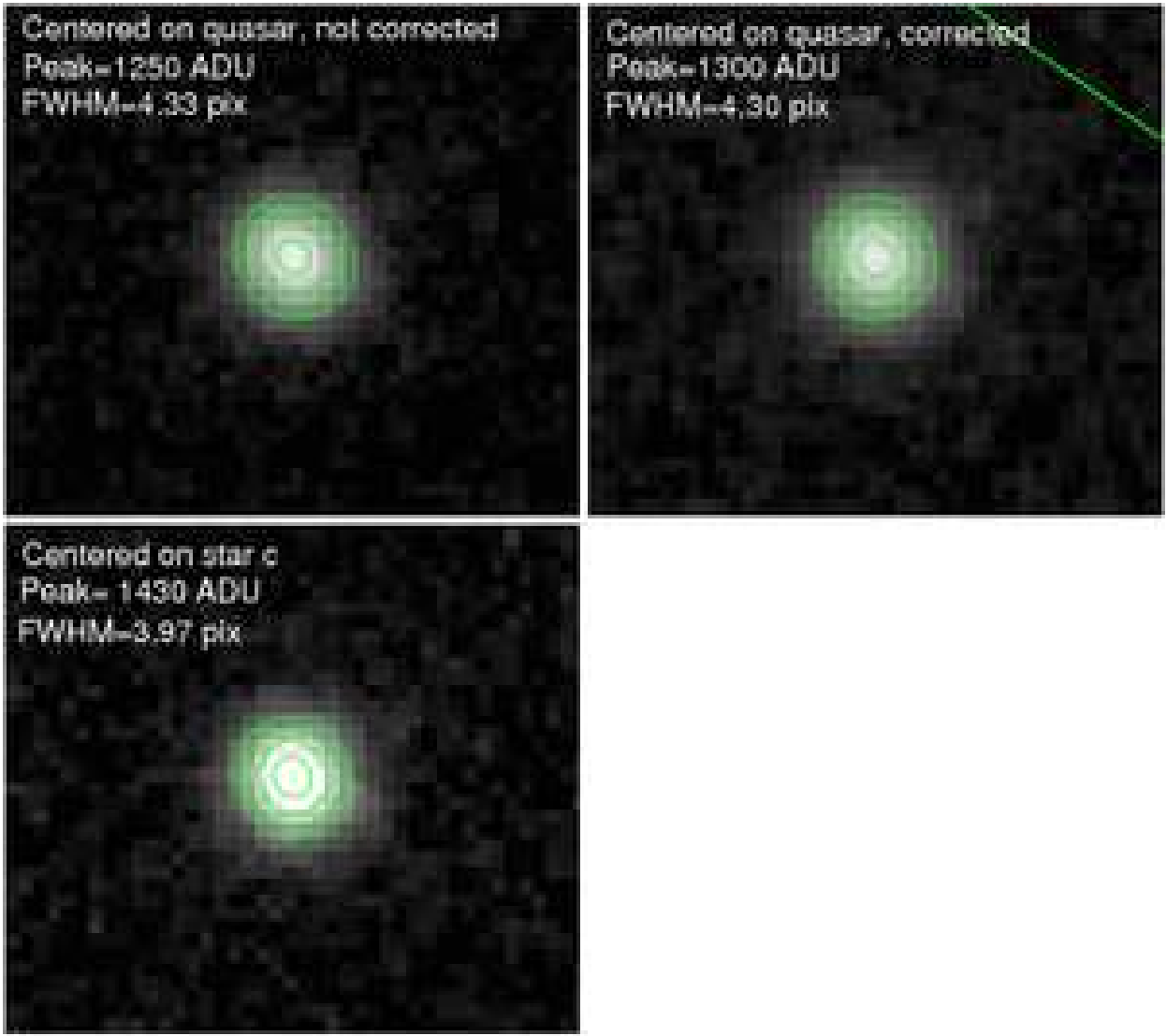}
\caption{Effects of distortion corrections and recentering for star ``c" from
  Fig. \ref{fig-bigimage}.  Top left: frames have been aligned to the
  quasar centroid before combining, and no distortion correction has
  been performed.  The star ``c'' image is stretched.  Top right: 
  the nominal distortion correction has been applied before aligning
  to the quasar centroid and combining, which has tightened 
  up the star ``c" image; however, it is still broader than the image
  of the quasar taken from the same frame. 
  Bottom left: the distortion-corrected
  frames have instead been aligned to the centroid of star ``c'' 
  before combining, effecting a second-order distortion correction for
  that star.  The star ``c'' image now has the same FWHM and shape as the
  quasar did in the image aligned on the quasar centroids.  \label{fig-distcontours}}   
\end{figure}

\begin{figure}
\plotone{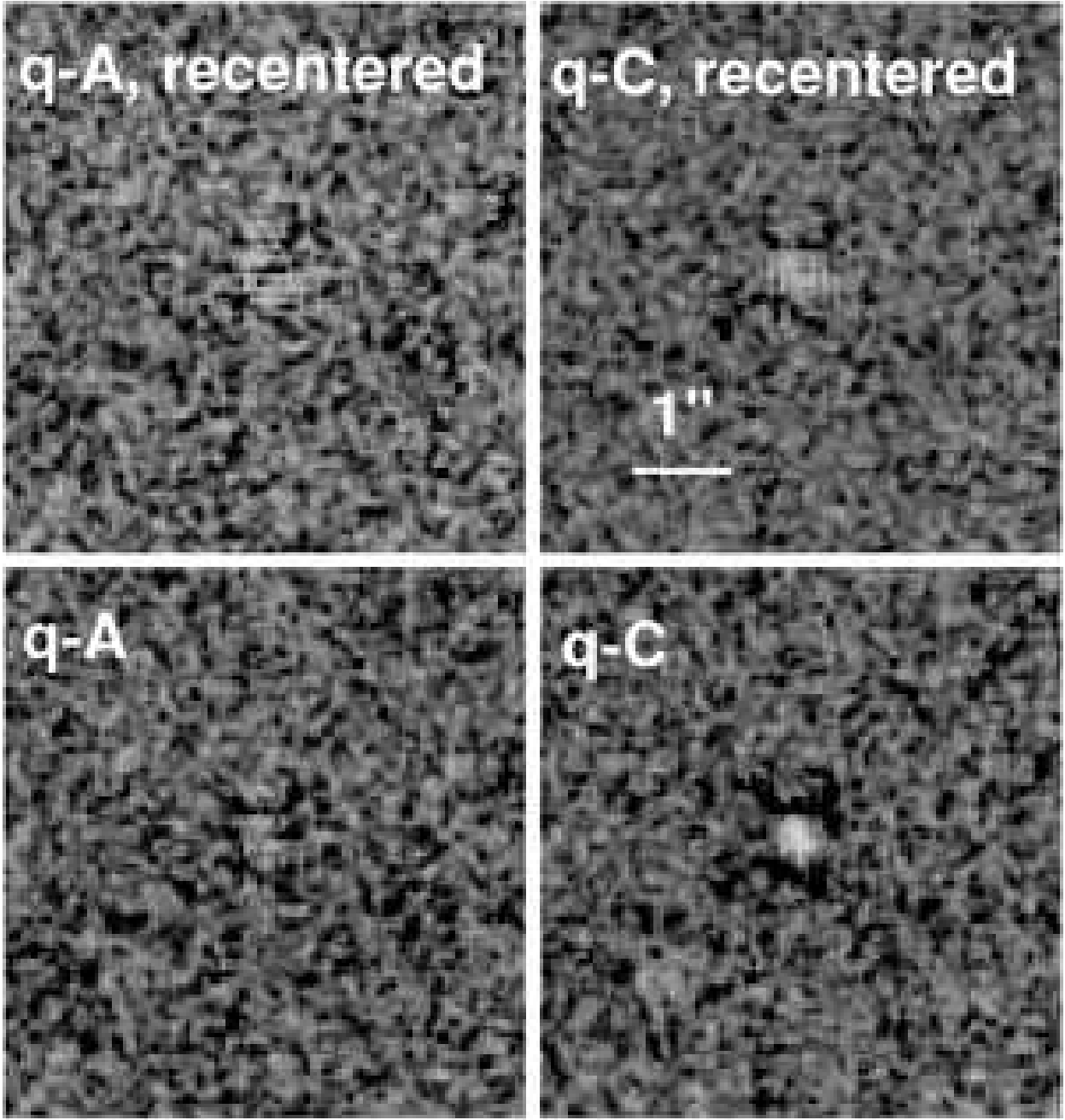}
\caption{Improvement of fits for the quasar shown in Fig. \ref{fig-bigimage} and \ref{fig-distcontours} after creation of ``recentered" PSF frames for stars ``a" and ``c."   The better fit with star ``a" is likely due to its relative proximity to the center of the frame.\label{fig-distfits}}
\end{figure}

To improve the fits, we were able to effect a higher-order distortion
correction by recentering each quasar's hundreds of frames on PSF
stars to create ``PSF frames" for each target, as suggested to us by
Brian McLeod.  With distorted images, the pixel scale at the edges is
different than that near the center.  Therefore, when shifting and
combining frames from different dither positions, the shifts computed
based on the objects near the center (in this case quasars) will be
the wrong number of pixels to align the objects near the edges,
yielding a combined image with stretched edges.  The recentering
technique to help correct for this works as follows. 
First, as described above, we align the hundreds of images to the
quasar centroids, and combine them to create a ``quasar frame.'' From
this we extract a postage-stamp image of the quasar to use for
fitting.   We then start again with the same hundreds of images and
align them  this time to the centroid of a particular PSF star, and
use these new shifts to combine them to make the ``PSF frame'' for
that star.  We then extracted a postage-stamp image of the PSF star
from the latter frame for use with the fitting.  An example is shown
in  Fig. \ref{fig-distcontours}.  We used this technique with good
success on most of the PANIC images.  Examples of fits performed with
and without our re-centering technique are shown in Fig. \ref{fig-distfits}. 

The NIRI images also suffer from distortion that proved significant
for this project.  No distortion correction is used in the NIRI
pipeline.  We applied our re-centering technique and found that it did
improve the NIRI fits considerably, but PSF-PSF tests (where we 
subtracted PSF stars from each other) showed that 
residual distortion remains in the K image of SDSSJ012019.99+000735.5 
and the H image of BRI0241-0146.  For these two images, 
the only PSF star is far from the quasar. 

The ClassicCam images provided their own set of challenges because the PSF
stars were observed alternately with the quasars, and inevitable
seeing variations resulted.  We were able to obtain more
reasonable results in most cases by rejecting frames according to the
seeing, so that the resulting quasar and PSF star frames had the same
FWHM.  However, the ClassicCam results are never as satisfactory or
robust as the Panic and NIRI results, and will be more useful for studies of the quasar's near environment than for host detection.

\subsection{Nonlinearity}\label{sec-nonlin}

Unfortunately, we also discovered that {\it the near-IR images exhibit
a small nonlinearity even after the nominal correction has been
applied}.  This  effect was more subtle than the distortion and became
apparent only under scrutiny of the ensemble of data for our many
objects.  Such an effect could easily have escaped our detection in a
study with 
fewer objects.  We noticed that our best fits were found for PSF stars
with similar brightness to the quasar.  PSF stars brighter or fainter
than the quasar could leave compact central emission or, more
insidiously, rings that mimicked host galaxies in the difference
images.    Star-minus-star experiments performed on multiple PSFs from
the same image confirmed our suspicion.   

\begin{figure}
\plotone{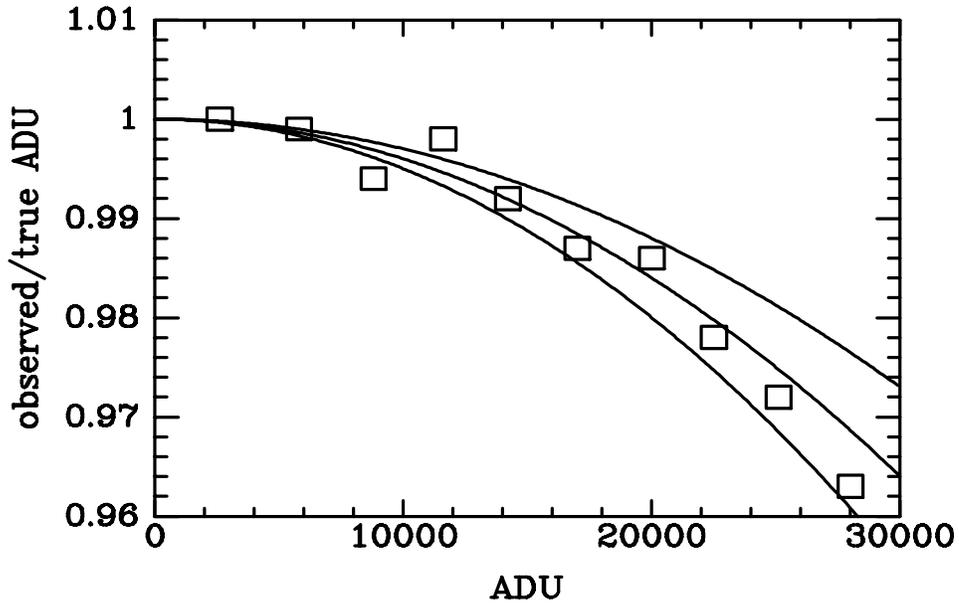}
\caption{Nonlinearity data for PANIC camera based on exposures of internal calibration lamps with three possible prescriptions for linearity corrections.  Exposure times are generally chosen to keep quasars (and PSF stars) below about 15,000 ADU.  At this level the possible corrections vary by $\lesssim 0.5\%$.  \label{fig-linplot}}
\end{figure}

\begin{figure}
\plotone{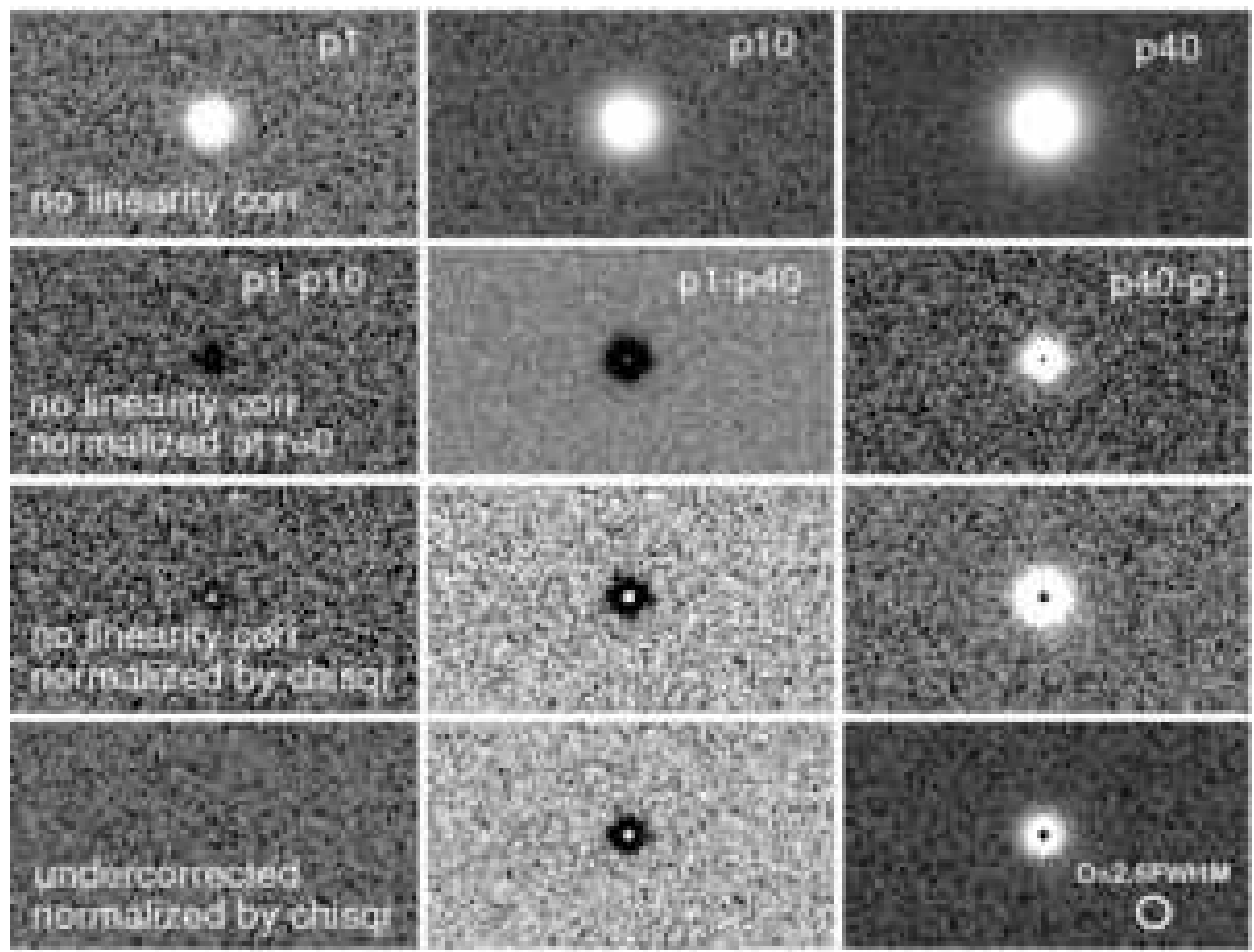}
\caption{Effects of imperfect linearity corrections on PSF fits.  Top row: simulated stars with the same FWHM and noise characteristics as our observed quasars and having relative flux 1, 10, and 40 times.  {\it Nonlinearity has intentionally been applied according to the middle curve from Fig. \ref{fig-linplot}.}  Bottom three rows: results of correcting and subtracting these stars from each other via different schemes.  In rows 2 (PSFs normalized to the central pixel before subtraction) and 3 (normalization based on a fit that minimizes residuals) we have performed no linearity correction.  Residuals are obvious and the details depend on the normalization.  In row 4, we have applied a linearity correction that differs from the one used to generate the stars by $\lesssim 0.5\%$.  The spurious residuals are only obvious with the brighter PSFs, and do not extend beyond a diameter of 2.5*FWHM.
\label{fig-linfigds9}}
\end{figure}

\begin{figure}
\plotone{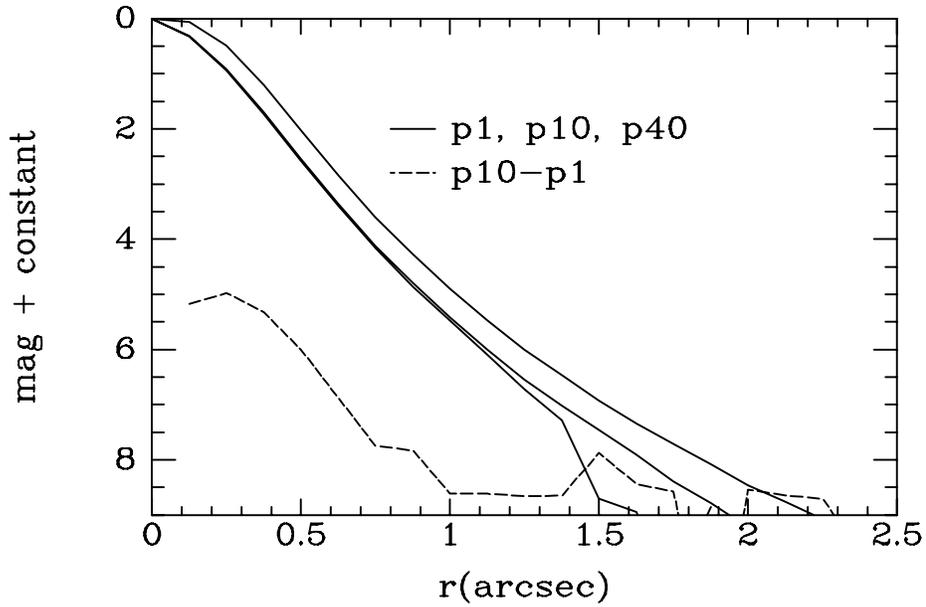}
\caption{Radial profiles for the simulated stars described in \S \ref{sec-nonlin} and shown in Fig. \ref{fig-linfigds9}.  The stars that have not been corrected for linearity.  For p40, whose flux puts it into the nonlinear regime, the profile is obviously different.  For p10 and p1, the more subtle nonlinearity falsely suggests a host contribution in the profile, though the residuals are unphysically compact. \label{fig-linfigprofs}}
\end{figure}

This is difficult to illustrate with observed stars because of the
residual distortion discussed above.  However, we investigated this
further by simulating observations of stars of different magnitudes
and fitting and subtracting them after applying different plausible
linearity corrections.  For example, we have used the nonlinearity
curve for PANIC shown in Fig. \ref{fig-linplot} to generate the suite
of stars shown in Fig. \ref{fig-linfigds9} with radial intensity
profiles given in Fig. \ref{fig-linfigprofs}.  The counts for the
fainter stars were chosen to keep the detector within the range where
the response is approximately linear, as is done for the observed
targets.  For the brightest stars, we allowed the brightness to enter
the nonlinear (but not close to saturated) regime.  For the PANIC
response, this corresponds to $\sim$15000 counts.  At this level, the
difference 
between the plausible prescriptions for the linearity correction amounts to
$\lesssim0.5\%$.  

When we generated stars using one prescription and then ``corrected"
them for nonlinearity using another, the 2D residuals in star-star
tests were clearly positive.  In other words, the uncertainty in the
linearity correction can lead to spurious detections when scaling and
subtracting point sources that differ in flux.  However, in the cases
we tried, the spurious residuals were distinguished either by
unphysically compact sizes (FWHM less than the image FWHM) as shown in
Fig. \ref{fig-linfigprofs}, or else by donuts that could be mistaken
for over-subtracted hosts.  The latter did not extend past a diameter
of $D=2.5~\rm FWHM$, as illustrated in Fig. \ref{fig-linfigds9}.  

\subsection{Implications}

Our results call into question the traditional approach of selecting
 ``PSF stars...chosen to be as bright as possible without encountering
detector saturation effects" \citep{hutchings2005}.  We have
used this approach ourselves for low-z quasars (e.g. \citet{mr95}) so that 
noise in the PSF wings is scaled down during the fitting process.  However, our
current analysis suggests that for PANIC and NIRI at least, a more
robust practice is to  chose PSF stars whose brightness is similar to
the quasar, and whose positions are as close as possible.  Which
criterion takes priority might depend on the instrument and the
observing conditions.   

Distortion corrections and linearity corrections are essential, but
not sufficient.  The PSF star re-centering technique described above 
provides a higher-order distortion correction, but a possible added complication is that the accuracy of the registration can be dependent on the brightness of the stars.
In addition, the characteristics of spurious residuals are dependent on
the weighting process used during normalization of the PSF
(e.g. normalize to the flux in the central few pixels, use the whole
source for the fit, weight the fit by flux, etc.).    

In principle, adaptive optics (AO) observations in which images of a
PSF are interleaved with those of the quasar should be free from {\it
geometrical} distortion when both are observed on the same part of the
array.  However, our results suggest that the case is not so clear.
First, the AO observations will suffer from the same nonlinearity
issues described above.  Second, the adaptive correction procedure can
be dependent on the object's flux and the details of the profile,
which can lead to an effective distortion.  This underscores the
desirability of observing PSF stars simultaneously with, and not just
close in time to, the quasar; see however \citet{ammons2009}. 

We conclude that the residual effects of distortion and nonlinearity
should be addressed by individual near-IR host-hunters for their
particular data sets.  For the present study, we evaluate the
residuals based partly on our knowledge of the brightness and
proximity of the PSF stars.  In most cases, we adopt as a criterion
that positive residuals are considered significant only if they extend
beyond a diameter of $D>2.5~\rm FWHM$, i.e. a radius  $r>1.25~\rm FWHM$, which
for the typical frame here means $r\gtrsim0.55\arcsec$.   We explore
this further with the simulations discussed below.  Of course, this
particular criterion might not be appropriate for data taken under
different seeing conditions or with different flux levels.   

\subsection{PSF Fits}\label{sec-fits}

To begin our search for host galaxies, we modeled each quasar as a
point source with shape represented by the PSF star images. 
We determined a two-dimensional (2D) best-fit model for each quasar
using the C program {\it imfitfits} provided by Brian McLeod and
described in 
\citet{leh00}.  This is the same program that we used on NICMOS images
of low-redshift quasars \citep{mm01}.  We used the 2x magnified images to
ensure good sampling of the PSF, and extracted an $8\arcsec\times8\arcsec$
sub-image for the fitting.  {\it Imfitfits} makes a model by convolving a
theoretical point source with the observed PSF, and then varying 
any combination of the parameters defining the background level and the 
position and magnitude of the point source to minimize the sum of the squares of the residuals over all the pixels.  
By subtracting the best-fit model from the quasar image, we can
examine the result 
for any residual flux due to an extended component.   

\begin{figure}
\plotone{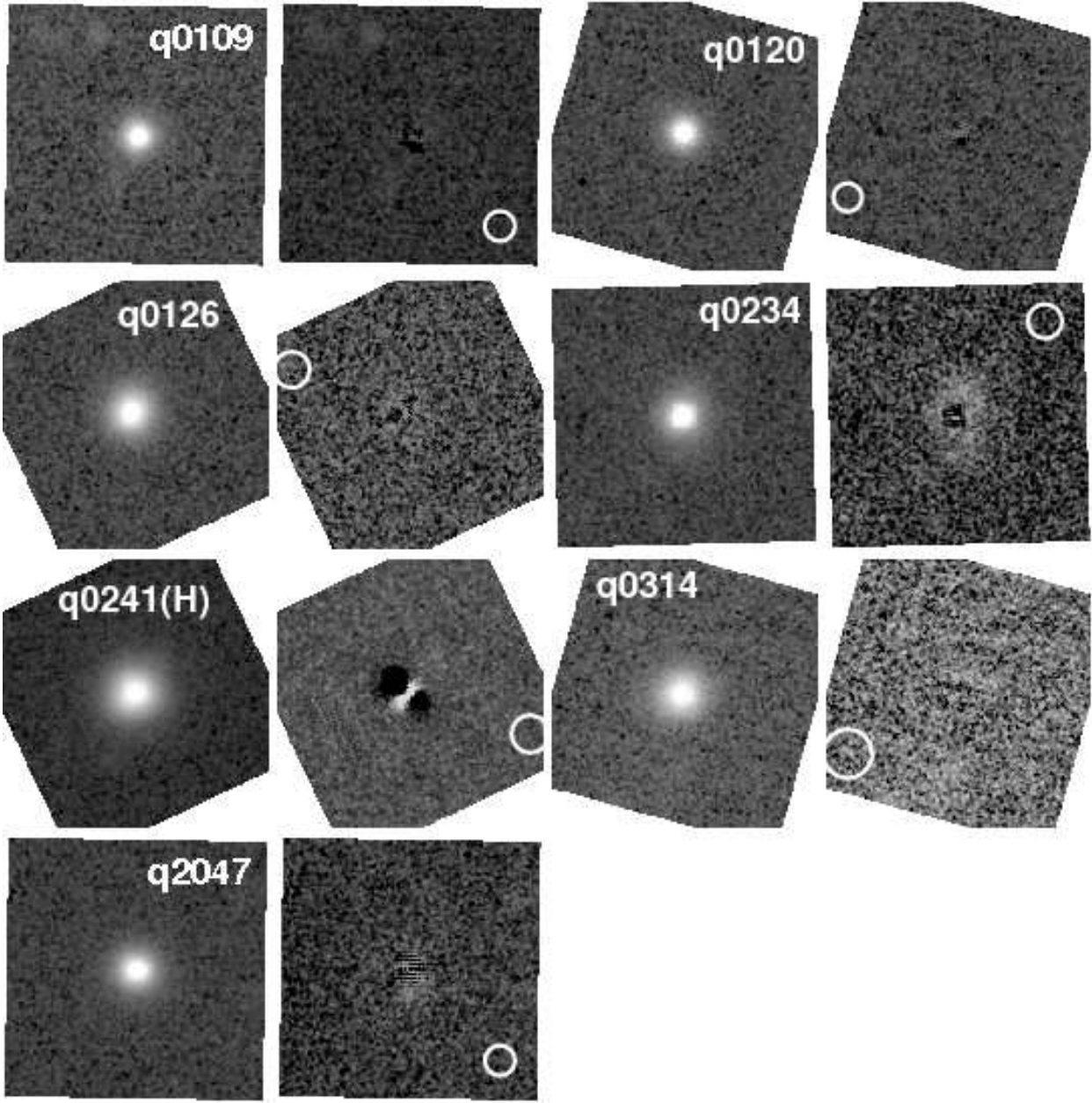}
\caption{Closeups of quasars (left image in each pair) and residuals after PSF fits, $\sim8\arcsec$ on a side for targets observed with NIRI.  Images are shown with North up, East to the left.  The circles have diameter $D=2.5\rm FWHM$.
The obviously bad fit for q0241H resulted from telescope mirror support problems, and illustrates the effects of distortion in an extreme case.  The q0120 image shows the much more subtle but typical effect of residual distortion and nonlinearity.
In the case of q0234, significant residuals are obvious.  
\label{fig-fits}}
\end{figure}

We achieved excellent results with at least one PSF star in most cases, as
shown in Fig. \ref{fig-fits}.  
In a few cases where other sources within the 8\arcsec\ box would
bias the fits of the quasar, we have simultaneously fit those other
sources as either point sources or galaxies.  In these cases, we
subtracted only the fitted quasar component for the figures; the other
sources remain for comparison. 

Our fitting process necessarily subtracts out any unresolved
contribution from the host.  A logical next step would be to perform a simultaneous fit of
a point source plus a model galaxy as is commonly done with lower-redshift quasars.
Unfortunately, we have found that for these data, multicomponent fits give uninterpretable
results.  The problem is that for our data, the likely range of scale lengths for the galaxies are small compared to the seeing disk and too little of the galaxy extends beyond the PSF.  The result is that running a multicomponent fit, whether unconstrained or partially constrained (for example by holding fixed the centers, or the centers and the galaxy shape, or the centers and the nuclear flux, ...), results in the "galaxy" component being turned into a meaningless compact or even negative source to improve the fit in the quasar's core, where PSF variations are the biggest.  One idea to get around this problem is to downweight or mask out the core in the fits.  However, our tests on real and 
simulated hosts have shown that the resulting "galaxy" is sensitively dependent on the weighting scheme.  (We had even seen hints of this with the much better resolved hosts in our low-z HST study \citep{mm01}.)  Thus, we have developed a different way to estimate host magnitudes and morphologies via the simulations and isophotal diameter analyses described below.   

\begin{figure}
\includegraphics[scale=0.8,angle=180]{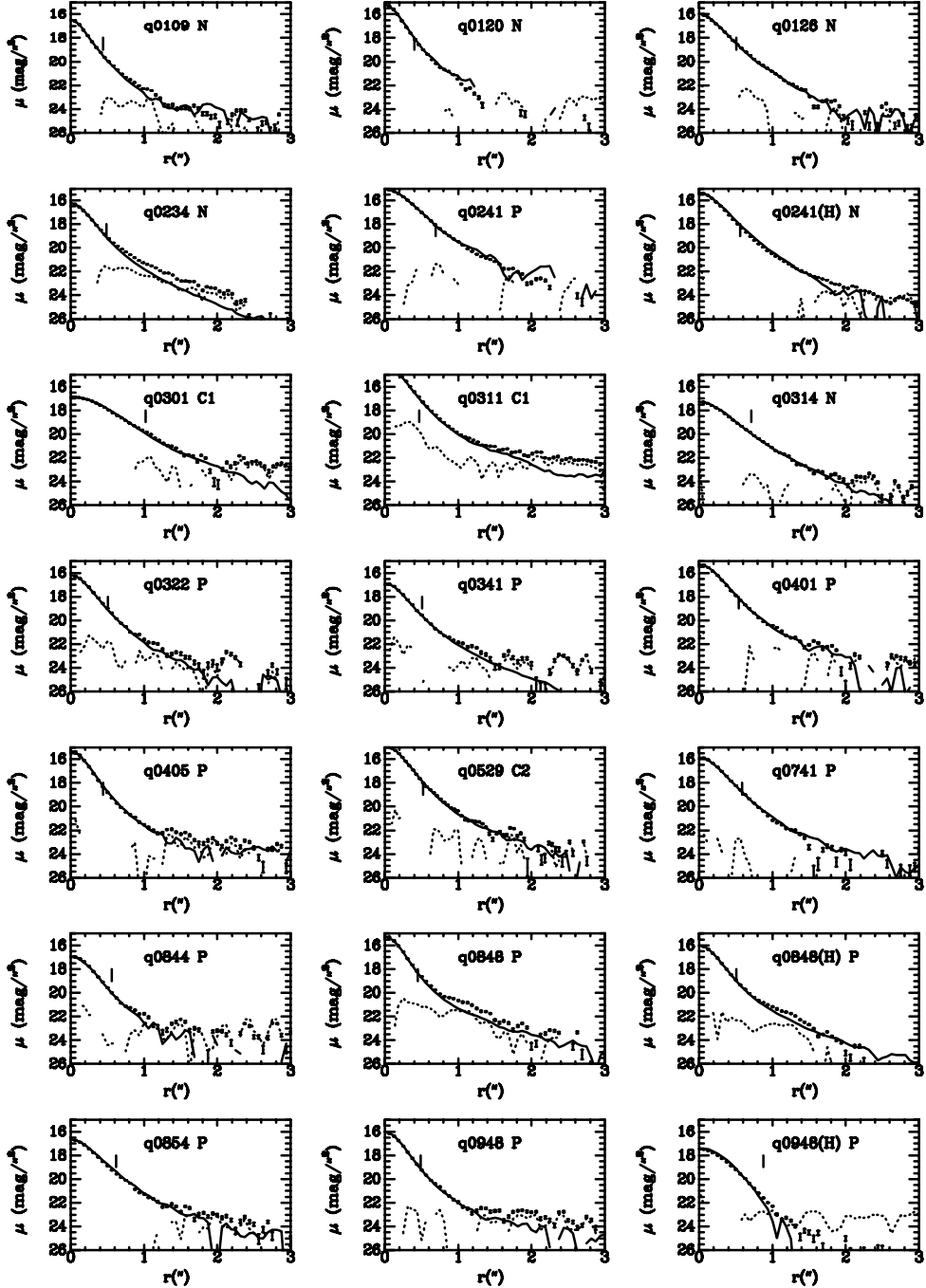}
\caption{Radial profiles labeled with instrument names (C1,C2=ClassicCam, N=NIRI, P=PANIC).  Quasar data points are plotted with standard-of-the-mean errors.  PSFs are shown as solid lines normalized to the central pixel.  Dotted lines show positive parts of residuals after subtraction.   Vertical bars mark the radius corresponding to a diameter as $D=2.5\rm FWHM$ discussed in the text.  \label{fig-profiles}}
\end{figure}

As another tool for assessing host detection, we have generated
one-dimensional (1D) radial brightness profiles, measured in circular
annuli, of the quasar and PSF images.   
We present the surface brightness profiles in Fig. \ref{fig-profiles}.
For comparison we also generated a ``fully-subtracted" profile by
normalizing the PSF to the quasar within the central few pixels and
subtracting. 
In most cases, the PSFs are excellent matches to the quasars down to
the level of the sky noise.  

We caution that the 1D profiles need to be interpreted carefully.  For
example, for our ClassicCam image of q0311-5537, the profile alone (see
Fig. \ref{fig-profiles}) looks like those of some host detections
postulated in the literature.  However, the 2D fit (see
Fig. \ref{fig-fits}) shows that the residual flux is due to PSF
mismatch.  For the cases where we do have candidate host galaxies, we
can use the 1D profiles to obtain estimates of the host flux.  To do
this, we subtract a fractional PSF profile that leaves a
just-monontonic residual (as any plausible host would not decrease in
brightness towards its center), and add up the residual light by
integrating.  This technique is necessarily crude, but the data do not
warrant more sophisticated fits.  

\section{Detection Limits}

In \S \ref{sec-psf} above, we concluded from our PSF-PSF tests that
any residual flux from our quasar fits outside a diameter of
$D>2.5~\rm FWHM$ is likely significant.  In this section, we explore
this criterion further in two ways: through simulations and through
calculations of predicted isophotal diameters for galaxies of various
types.  This second technique is (as far as we know) a new  and
potentially very useful approach for quasar host galaxy studies. 

\subsection{Simulated Hosts}\label{sec-sims}

To probe our detection limits we generated suites of fake galaxies and
added them to the magnified images of apparently point-like quasars.
We selected quasars both brighter and fainter than the median for our
sample, and images both at, and deeper than, the median surface
brightness limit.  We convolved each model galaxy with the quasar
image, added the result to the quasar image itself, and reran the
analyses with the PSF stars.  We also duplicated some of the tests
with noiseless (Moffat) PSFs having the same FHWM as the quasars.  Our
simulated galaxies included both exponentials (central surface
brightness $\mu_0$, scale length $r_0$) and deVaucouleurs profiles 
(surface brightness $\mu_{eff}$ at effective radius $r_{eff}$).  We
considered sizes $r_0$,$r_{eff}$ of 0.125, 0.25, 0.5, and 0.75\arcsec,
corresponding to 0.88, 1.5, 3.5, and 5.2kpc at $z=4$.  These values
are similar to
the range observed for $z\sim4$ galaxies in the Hubble Ultra-Deep Field (HUDF)
\citep{elmegreen2007} and high-z lensed quasar hosts \citep{peng2006}.
We tested axial ratios $0.2<b/a< 1$.   

Visual inspection of the residuals supported the validity of our
$D>2.5~\rm FWHM$ criterion; detectable galaxies left residual light outside of that diameter.  In terms of flux, we found that for the
galaxies we tried, the hosts were cleanly visible for an observed $\rm
K_s-band$ flux ratio $F(host)\gtrsim \onethird F(nucleus)$.  These
hosts leave central (negative) holes in the subtracted 2D images and
have flux in clear excess of the PSF at $r\approx1\arcsec$ in the 1D
profiles.  For hosts with $F(host)< \onethird F(nucleus)$, the detectability by visual inspection
depends on the size.  The hardest galaxies to recover were those whose
scale lengths or effective radii were $\rm <\onethird FWHM$, and also
the very large $r_{eff}=0.75\arcsec$ deVaucouleurs galaxies for which
too much of the galaxy's flux is at low surface brightness.   

\subsection{Isophotal Diameter Analysis}\label{sec-isod}

Bolstered by the results from our simulation, we recast our detection
criterion from Section \ref{sec-psf} above as a detection limit in terms of
galaxy isophotal diameter $D_{iso}$, here taken to mean the diameter at which the galaxy light drops below the sky noise.  In other words, we assume that we can detect any hosts that have $D_{iso}\gtrsim2.5~\rm FWHM$ for the surface brighteness limits of our images.  

To explore the range of galaxies that could be detectable as hosts, we have calculated $D_{iso}$ for model exponential and deVaucouleur galaxies following the tradition of \citet{weedman86} but updated for the currently favored cosmology.
We start with exponential and deVaucouleurs galaxies covering a range of scale lengths similar to those in \S\ref{sec-sims}.  We transport them to $z=4$ by applying  cosmological surface-brightness dimming and cosmological angular diameter distances. 
We calculate their $z=4$ colors and k-corrections by redshifting and integrating a spiral galaxy spectral energy distribution template over the filter bandpasses.  
[We have also used bluer and redder templates, but we note that for these data, the k-correction is nearly independent of galaxy spectral shape because the observed $K_s$-band corresponds to rest-frame $B$.]  Finally, we combine these to calculate the isophotal diameters for the galaxies given the surface brightness limits of our images.  
We also compute each galaxy's observed magnitude $m_{Ks}\rm (obs)$ by integrating the galaxy flux inside the isophotal diameter.

Fig. \ref{fig-isod}a shows $D_{iso}$ as a function of observed
magnitude $m_{Ks}\rm (obs)$ for a range of galaxy types and sizes, and
for the range of surface brightness limits found for the 2x magnified
PANIC and NIRI images that we use for host detection.  
We stress that $m_{Ks}\rm (obs)$ represents only the fraction of the galaxy's light that
falls above the sky noise; it is not simply the absolute magnitude adjusted by the cosmological distance modulus.   
On these plots, only galaxies above the $\rm 2.5 FWHM$ line are in
principle detectable as hosts. 
One can see that for the galaxies considered (i) the faintest visible
deVaucouleurs hosts span $\sim 1mag$ at a given surface brightness
limit;  (ii) the faintest visible exponential hosts span $\sim
1.5mag$; and (iii) deVaucouleurs hosts must be relatively brighter to
be detected because more of their light is hidden in the steeply
sloped and unresolved core. 

In Fig. \ref{fig-isod}b, we plot these $D_{iso}$ values as a function
of \lstar~ assuming {\it no evolution} in the mass-to-light ratio of the stellar
population.  We adopt for reference a local
\lstar\  galaxy of magnitude $M_B^{*Vega}=-20.5$.  Transported to
$z=4$ such a galaxy would have a total magnitude of $m_{Ks,\rm
no~evolution}^{*Vega}=23.6$, whereas its observed magnitude would be
fainter depending on the surface brightness limit.  One can see from
the figure how the detectability depends on the galaxy scale length.
For example, on an image with the median FWHM and with the deepest
limiting surface brightness ($\mu=22.4\msa$), the intermediate-scale
($r_0=1.5\rm kpc$) exponentials are visible at lower luminosity than
either the small- or large-scale exponentials.  The smaller galaxies
hide a larger fraction of their flux in the unresolved core.
The larger galaxies have lower central surface brightnesses at
a given luminosity, and their relatively more shallow disks do not pop
above the surface brightness limit until farther out in their
profiles.   
We use these plots to estimate host detection limits for
each object.   

We note that there are different ways that one might measure the surface-brightness limit of any given image.  We have compared our calculated isophotal diameters and apparent magnitudes from this section 
with those measured on the images for the simulated galaxies discussed above in \S\ref{sec-sims}.  We find them to be in excellent agreement when the surface brightness limit used for the isophotal diameter calculation is that given by the $1\sigma$
pixel-to-pixel sky noise.  This is how we have characterized the
surface brightness limits for our images in Table \ref{tab-obs}. 

\section{Results}

In a typical image, we are sensitive to {\it field} galaxies as faint
as $m_K^{Vega}\sim23$  ($m_K^{AB}=24.8$), with the actual limits
dependent upon morphology. To translate this apparent magnitude into 
a corresponding luminosity for a present-day galaxy with the same stellar mass, 
we need to account for luminosity evolution of the stellar population.
A reasonable assumption for the
evolution is that the galaxies undergo $\rm \sim2~mag$ (i.e a factor of 6) of fading between $z=4$ and now,
which we infer from the $K$-band k-corrections measured for galaxies in the HDF-S \citep{saracco06}.
This amount of evolution is also expected in the (rest frame) visible mass-to-light ratio according to the stellar population synthesis models of \citet{bruzual2003} for formation redshifts of $z\gtrsim5$.
If this is the case, our 
images yield galaxies with stellar masses corresponding to a present-day galaxy with luminosity 
$\lesssim \lstar$ in the fields around the quasars.  A detailed study of the quasar environments will be presented in Bechtold \& McLeod (2009).  Here we discuss the results
for hosts. 

\subsection{Host Limits}

\begin{figure}
\plotone{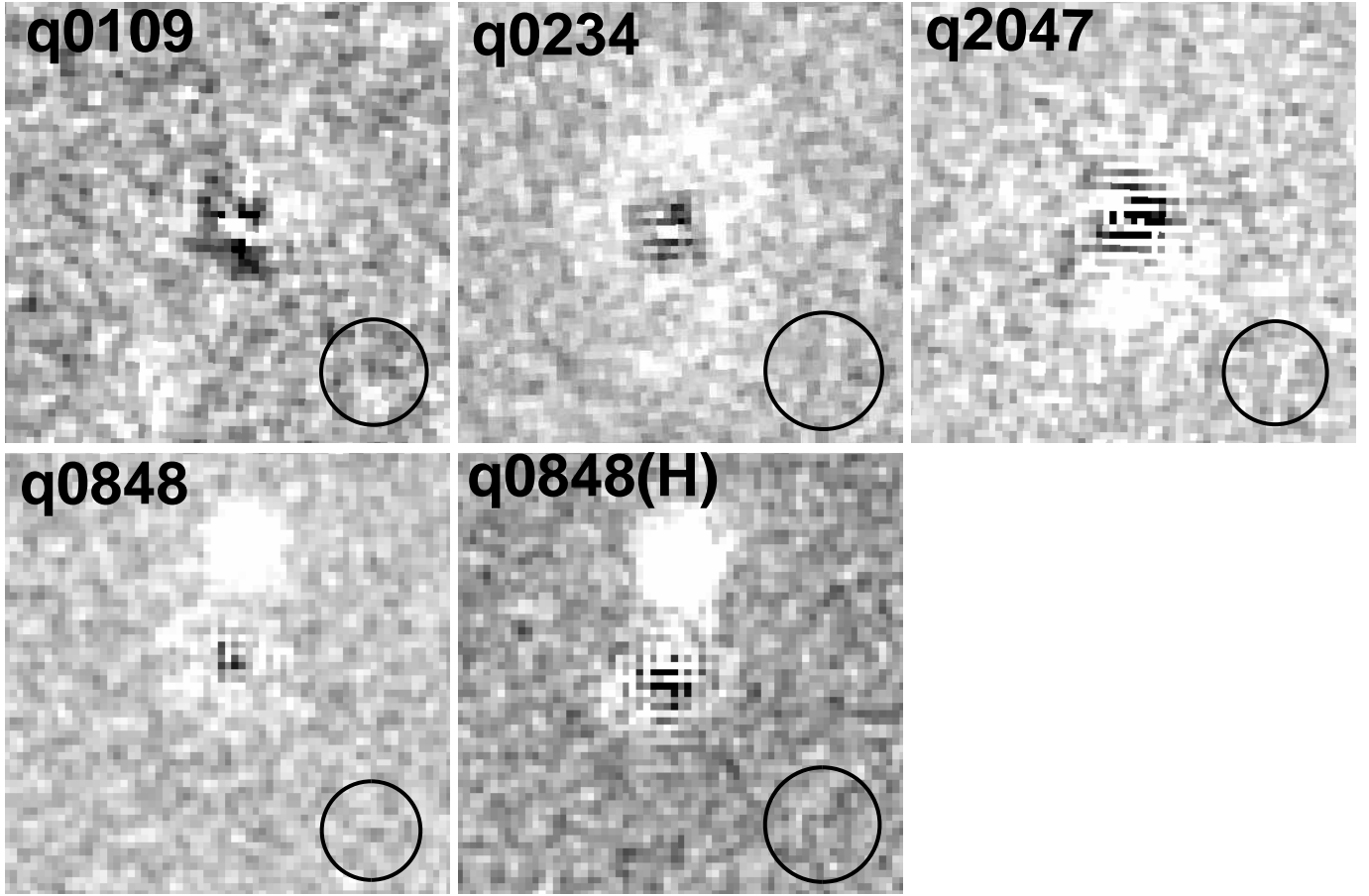}
\caption{Closeups of likely hosts.  Boxes are 4\arcsec on a side with North up, East left, and circles as in Fig. \ref{fig-fits}.  Top row: NIRI targets.  Bottom row: PANIC object q0848 in K and H. \label{fig-fourhosts}}
\end{figure}

For {\it host} galaxies, our detection limits are of course brighter than
the limits for galaxies in the field.
We inspected the radial profiles together with the two-dimensional
fits to classify detections as y/?/n (likely/maybe/unlikely) with
results given in Table \ref{tab-hosts}.  We looked for residuals that
extend beyond the sizes of the circles shown in Fig. \ref{fig-fits},
and that are not likely attributable to nearby (projected on the sky) companions.
We further used the 1D fits to verify that the residuals were
plausibly broader than the PSF.  {\it We find four likely
hosts, and note that they are seen on the images that have the best seeing, $\rm
FWHM<0\farcs4$, and nearly maximal depth, indicating that we are
pushing the limits of detection with these images. Other
objects might well have hosts lurking just beneath the noise.}  Three
of the four likely detections are found on NIRI images.  The fuzz
associated with quasar q0848 in the Ks band was marginally detected in
H as well.  These four likely hosts are shown in Fig. \ref{fig-fourhosts}.

For all of the objects observed with PANIC or NIRI, we estimate the
host detection limits by applying our $D_{iso}\gtrsim2.5~\rm FWHM$
criterion using the curves in Fig. \ref{fig-isod} and the surface
brightnesses and FWHMs in Table \ref{tab-obs}.  The results are
summarized in Table \ref{tab-hosts}, where we list for each model two
possible values for the limit on the host galaxy.   

The ``conservative" value represents the most luminous host that would
be just visible; the galaxies may be luminous but are conspiring to
evade detection either by putting too much light in their unresolved
cores or by having such a large scale length that the middle radii are
below the sky. 
The ``optimistic" value represents the least luminous host that would
be just visible. 
Of course the stellar masses of the galaxies in Table
\ref{tab-hosts} could be considerably smaller than the straight luminosities
indicate. For example, if we allow for 2 mag of evolution, then the
the present-day equivalents would be galaxies lower in luminosity by a
factor of $\sim 6$.  In that case, a 12\lstar\ galaxy in the Table
 would represent 2\lstar\ of stellar mass. 

One can see from Table \ref{tab-hosts} that for the depth and resolution
of our images, the range in upper limits for each type of galaxy
typically spans a factor of two.  In addition, the luminosity limits for
deVaucouleurs galaxies are typically double those for exponentials,
reflecting the fact that the peakier spheroids can hide more light in
the unresolved core.  In general, the least certain limits by the
isophotal diameter method will be for those images with big FWHM
and/or shallow depths, because in these cases galaxy isophotal
diameters are only weakly dependent on galaxy luminosity--in other
words, they fall on the flat outer parts of the curves in
Fig. \ref{fig-isod}b.  The ClassicCam images were sufficiently
insensitive that the limits are not interesting.  There are also
several PANIC and NIRI images whose conservative limits were so large
as to be also uninteresting. 

\subsection{Host Detections}

\begin{figure}
\epsscale{1.1}
\plotone{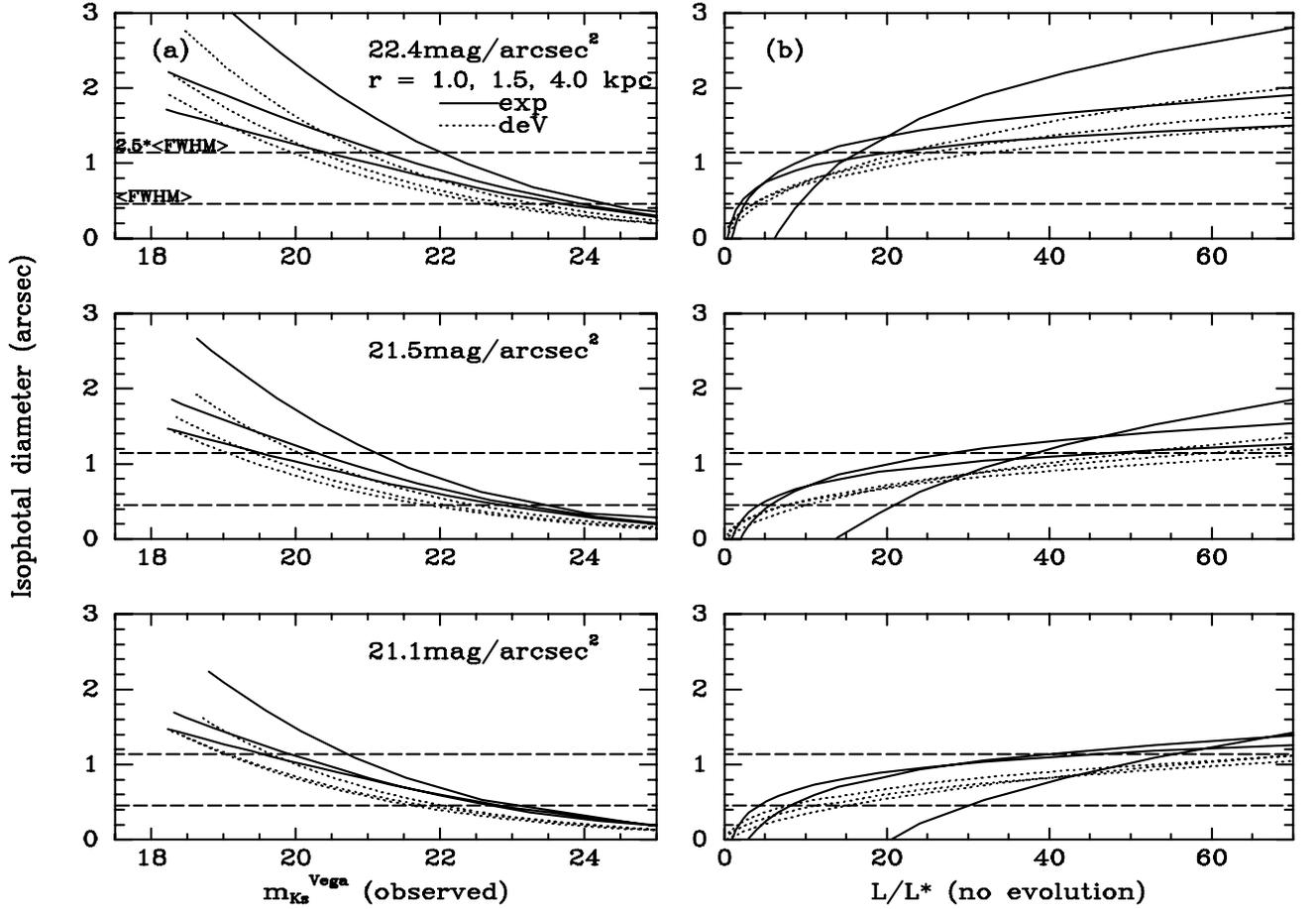}\caption{Isophotal diameter predictions v. observed magnitude $m_{K_s}^{Vega} \rm (obs)$ (a) and luminosity $L/L^*$ (b) for model galaxies at $z=4$. 
The observed magnitudes include only the galaxy flux that is inside the isophotal diameter.  
The luminosity plots assume no evolution; with a more realistic 2 magnitudes of evolution the luminosity axis labels will decrease by a factor of $\sim6$. 
The quantity $r$ represents $r_0$ and $r_{eff}$ for exponential and deVaucouleurs laws respectively.  The three vertical panels in each set represent surface brightness limits appropriate for the range of our NIRI/PANIC observations.  Dashed horizontal lines show the median FWHM and $2.5~\rm FWHM$.  As described in \S \ref{sec-isod}, we set detection limits for the various host galaxy types according to the criterion $D_{iso}\gtrsim2.5~\rm FWHM$.
\label{fig-isod}}
\end{figure}

For the four likely hosts we can also estimate fluxes from the
residuals after fitting.  We use Fig. \ref{fig-isod}a first to identify
the kind of galaxy that could yield the {\it observed} magnitude and
isophotal diameter for the surface brightness limit of the image.  We
then use Fig. \ref{fig-isod}b to translate that into a possible intrinsic
B-band luminosity for the whole galaxy.  Note that this luminosity is
bigger than the luminosity one would calculate simply from applying
the distance modulus to the observed magnitude; it includes
contributions from the inner part of the galaxy under the PSF and from
the outer part below the sky noise.  Finally, we apply 2 mag of
evolution to the luminosity and from it calculate the corresponding
$M_B$ that a galaxy of the same stellar mass would have today. 

For q0109, the residuals add up to $m_K^{Vega}\rm (obs)\sim22.7$ and
they extend to a diameter of approximately 2\farcs.  We
use Fig. \ref{fig-isod}a (the 22.4\msa~depth is appropriate for this
image) to learn that this galaxy could for example be a large 
scale-length exponential disk.  Locating this curve on
Fig. \ref{fig-isod}b, we find that the same $D_{iso}$ gives a luminosity
of $\sim 30\lstar$ with no evolution.  Allowing for 2 mag of evolution
yields a galaxy with mass corresponding to a $\sim5\lstar$ galaxy
today.   
For comparison, the object at $1\farcs5$ southeast of the quasar
has $m_K^{Vega}\rm (obs)\sim22.7$ and $D_{iso}\sim 1\farcs1$.  
If it is at the redshift of the quasar, it could represent a companion with
roughly half the mass of the host at a projected separation of about 11kpc.

For the bright residuals in q0234, we measure $D_{iso}\sim 2\farcs6$,
while integrating the 1D residuals gives $m_K^{Vega}\rm
(obs)\sim19.9$--bright enough to represent a $\sim60\lstar$
exponential (10\lstar with evolution), again with a large scale
length.  

A similar analysis for q0848 gives $D_{iso}\sim1\farcs2$ and
$m_K^{Vega}\rm (obs)\sim19.4$ which could be a $\sim24\lstar$
(4\lstar with evolution) disk.   

For q2047, the residuals extend asymmetrically to the south and
possibly represent the combined flux of the host and a companion.
This object's classification as a hyperluminous infrared galaxy by \cite{rowan2000} 
(based on sub-mm emission that is likely too strong to originate in
the quasar's dust torus)  
would suggest that we are observing a merger.
Integrating the 1D residuals gives $m_K^{Vega}\rm (obs)\sim19.7$.
The diameter is harder to define because of the asymmetry but we adopt
$1\farcs6$ as an estimate, which implies a $\sim40\lstar$-mass,
intermediate scale length exponential (7\lstar with evolution).

\begin{figure}
\epsscale{1.7}
\plottwo{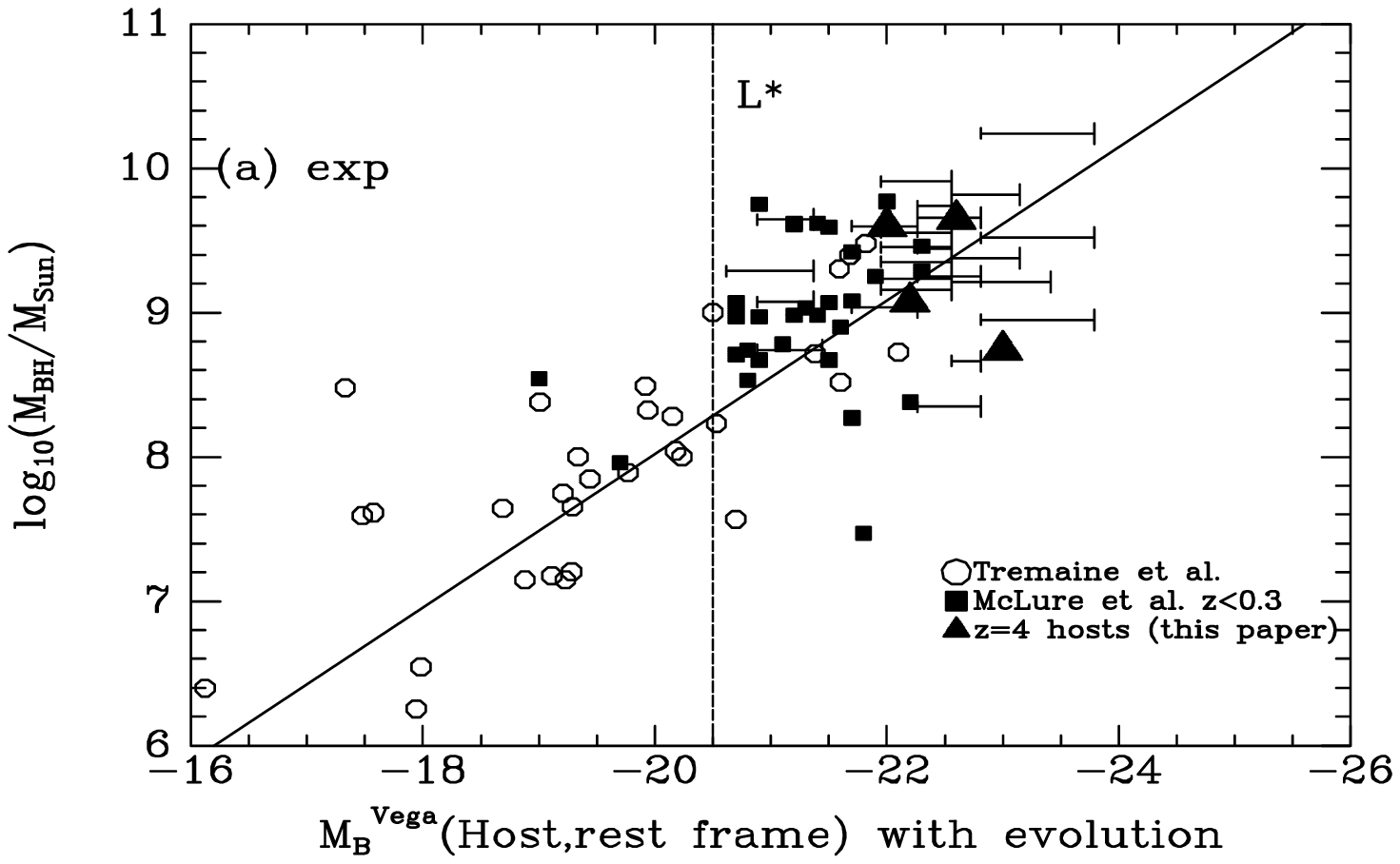}{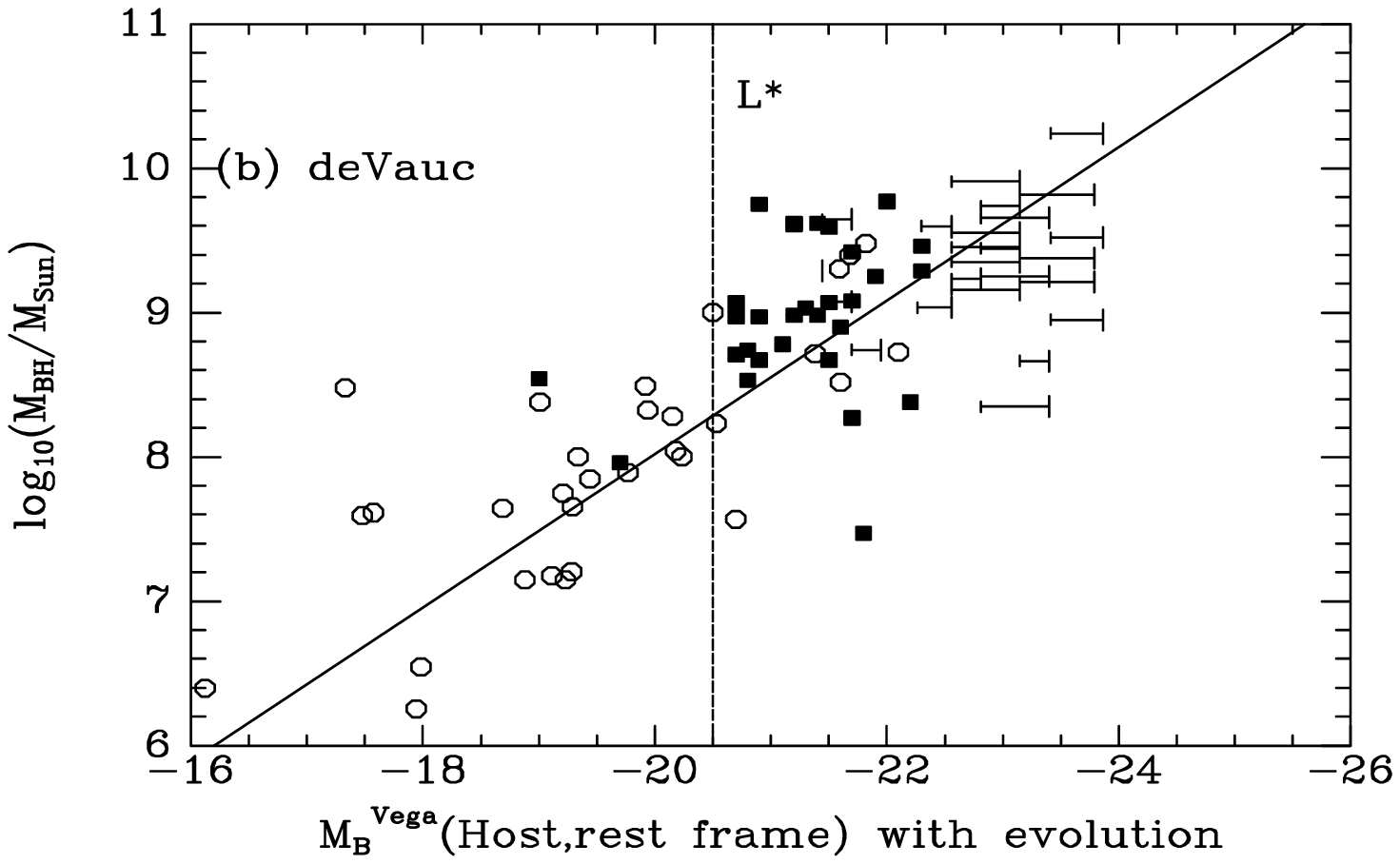}\caption{Comparison with the
  local black-hole/bulge relation.  Local galaxies are represented by
  the open circles from \cite{tremaine2002} plus the diagonal line fit
  from \cite{lauer2007}. Local luminous quasars from \cite{mclure2001}
  are shown as filled squares.   For each of our objects, the two
  connected vertical bars mark the optimistic (left end) and
  conservative (right end) upper {\it limits} on luminosity for the
  range of galaxy   scale lengths represented in Fig. \ref{fig-isod}.
  Limits are plotted for (a) exponential hosts and (b) deVaucouleurs
  hosts.  Triangles indicate the host luminosity estimates for the 4
  host galaxies detected.  Our host magnitudes and limits have been
  corrected for 2 mag of evolution.  Data from ClassicCam are not plotted, for
clarity.  All values from the literature
  have been adjusted to \Hseventy.   \label{fig-lauerline}} 
\end{figure}

Table \ref{tab-hosts} shows that three of the four detections are more
luminous than the conservative detection thresholds for exponentials and all
fall above the optimistic thresholds.  We overplot the estimates on 
Fig. \ref{fig-lauerline}-\ref{fig-allmags}.

If we examine the deVaucouleurs curves in Fig. \ref{fig-isod} we find
that the residuals for q0109 and q0234 are too big for their
magnitudes observed to be represented by the curves plotted; in other
words, if they are spheroids, they must have $r_{eff}>>4\rm kpc$.   On
the other hand, the residuals for q0848 and q2047 fall on the curves
for $r_{eff}=4\rm kpc$ and could be spheroids with masses of 1-2 times
their exponential counterparts.

For the several quasars with detections listed as ``?" the data do not
warrant any attempt to characterize the magnitudes other than to say
that if the hosts are there, they likely lie close to the limits
listed in Table \ref{tab-hosts}. 

\subsection{Color of the Host Galaxy of SDSSpJ084811.52-001418.0}

For one of the quasars with a detected host, SDSSpJ084811.52-001418.0, 
we also obtained a deep H-band image with PANIC.  We carried out the PSF 
estimation and subtraction from the nuclear quasar image for the H-band,
and found a residual flux, consistent with the K-band
detection.  The color of the host light is H-K=0.8.  This color is 
nominally bluer than that of a redshifted spiral galaxy spectral energy 
distribution, which would have H-K=2.0.
The quasar itself has H-K=0.5, 
as expected for a quasar at this redshift \citep{chiu2007}.   

Thus, the host galaxy is redder than the quasar itself, but bluer than
a young stellar population at $z=4$.  It could be that the host is 
experiencing a burst of star-formation, as one would expect to occur for
a major merger of gas-rich galaxies, and is metal-poor, so has a weak 
4000\AA break.  Another possible explanation, however, is that 
the host light is contaminated by a foreground galaxy.   There is no   
known damped Ly-$\alpha$ absorber along the SDSSpJ084811.52-001418.0
sight-line \citep{murphy2004} but other intervening absorption line 
systems are no doubt present. Finally,  we note that the uncertainty in
the H-K color is very large, and difficult to estimate.  
Deeper imaging of a larger sample, or spectroscopy of the fuzz, could
distinguish among these possibilities.

\subsection{The Local Black-hole/Bulge Relation}

In this section we compare our limits with the local black-hole/bulge
relation.  While the local relation is given for spheroids, we include
both exponential and deVaucouleurs models in our discussion to allow
for the possibility that the stellar mass might be differently
distributed at early cosmological times.  Wherever necessary we have
adjusted values to our adopted cosmology, and we have transformed
host absolute magnitudes in the literature to
$M_B$ assuming galaxy rest-frame colors $B-V$=0.8, $B-R=1.4$,
and $B-H=4$, all appropriate for a spiral-like stellar population.
For an older (elliptical-like) population the colors are up to 0.3mag
redder, but the uncertainties in these colors are small compared to
the other uncertainties.  

We have computed rest-frame absolute $B$ magnitude limits for our hosts 
from the luminosities in Table \ref{tab-hosts} assuming 2 magnitudes of 
evolution.  We plot our host limits and black hole masses against the 
local relation in Fig. \ref{fig-lauerline}, where the local population 
is represented by (i) the \cite{tremaine2002} local galaxies, (ii) 
the \cite{lauer2007} local fit, and (iii) the \cite{mclure2001} local 
($z<0.3$) luminous quasars, the latter which provide high-mass black hole 
counterparts analogous to those in our sample of high-z quasars.  The 
results are similar if we use our $0.4 L_{edd} $ estimates for the
black hole mass.  We interpret this figure as follows.  The conservative 
case occurs where the hosts are all at their maximum allowed values, i.e. 
with limit given by the right-hand bar of each pair and hiding much flux 
in either a very compact core or in the low surface brightness wings of a 
shallow profile.  In this case, the exponential and deVaucouleurs 
distributions are both consistent with the local relation for the two 
magnitudes of evolution assumed; the limits thus do not provide interesting 
constraints on possible evolution in the relation.
The optimistic case occurs when the left-hand bar in each pair gives 
the {\it upper limit} to the galaxy luminosity.  In this case, the limits 
for deVaucouleurs hosts remain uninteresting. On the other hand, hosts 
with exponential profiles would be less massive for a given black hole 
mass than are the local (Tremaine) galaxies, though might yet be consistent 
with local quasars.   However, {\it if the evolution correction are 
more than the two magnitudes assumed, our optimistic exponential upper 
limits would yield hosts lower in mass for a given black hole mass than 
local luminous quasars.}  The \citet{bruzual2003} models suggest that 
evolution in excess of 2 magnitudes between $z=4$ and now would be 
expected if for example the population were younger than about 400Myr 
at $z=4$.  



\begin{figure}
\epsscale{1.7}
\plottwo{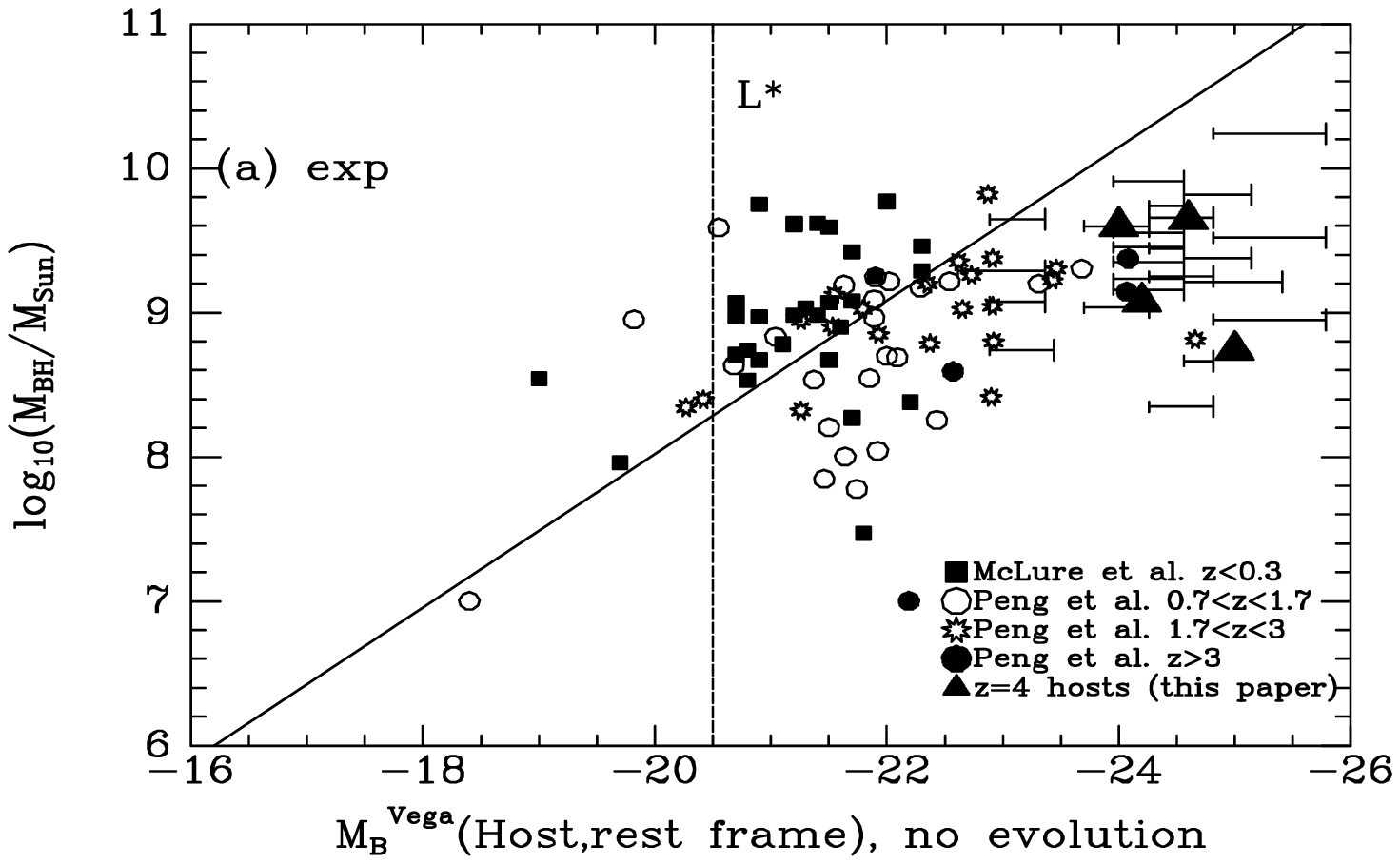}{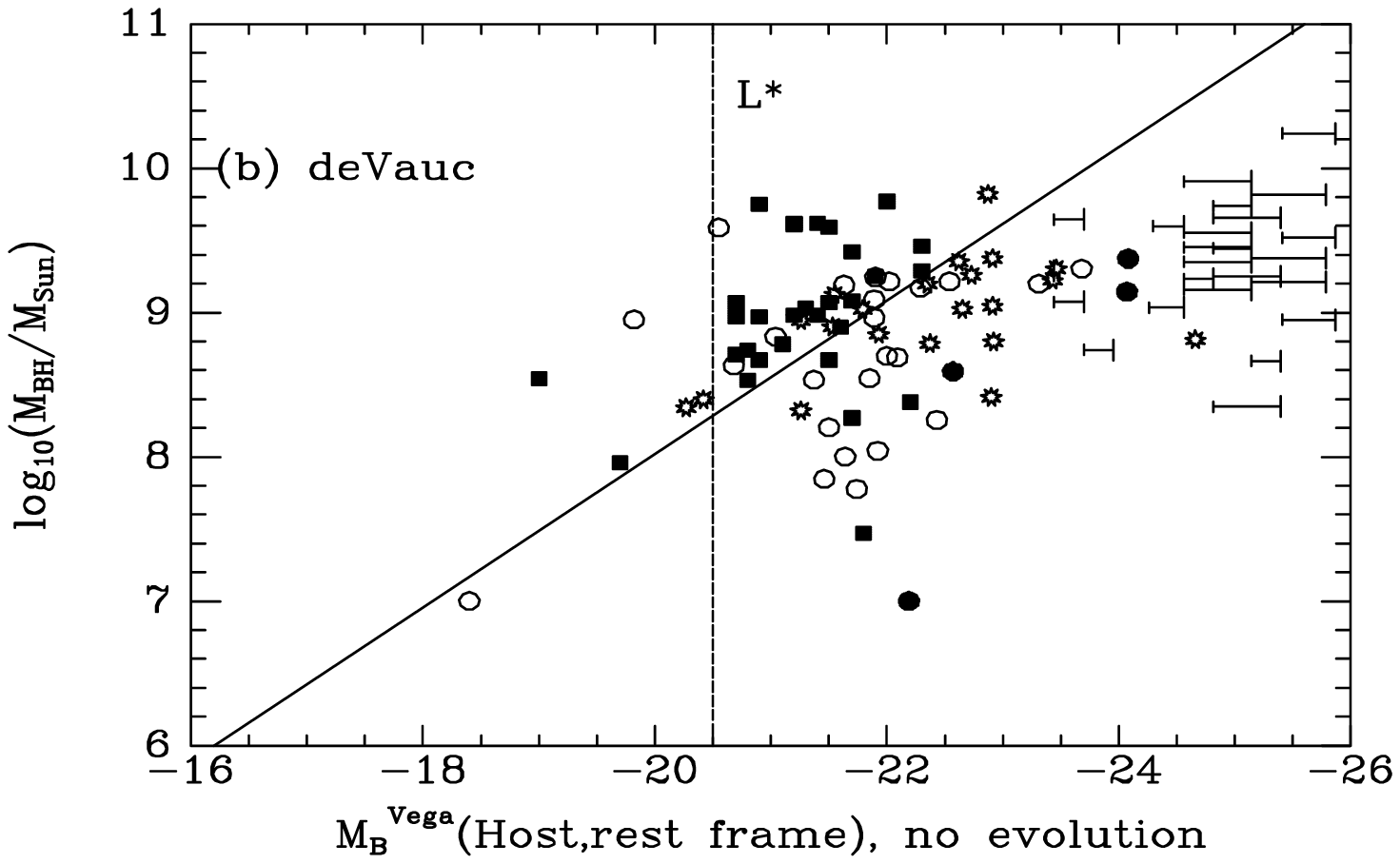}\caption{Comparison with quasars at
  different redshifts.  Local luminous quasars (filled squares) and
  the diagonal line are the same as in Fig. \ref{fig-lauerline}.
  Intermediate redshift quasars from the \cite{peng2006} compilation
  are shown as circles.  For each of our objects, limits are plotted
  as in Fig. \ref{fig-lauerline} except that here we have made {\it no
    correction for evolution}. Data from ClassicCam are not plotted, for 
clarity. \label{fig-peng}} 
\end{figure}

Fig. \ref{fig-peng} shows a similar comparison for quasars at various redshifts.  The local quasar sample is the \cite{mclure2001} local ($z<0.3$) used above.  Intermediate-redshift objects ($z>1$) are from the  \citet{peng2006} compilation of quasars observed with HST's NICMOS and having virial black hole mass estimates.  They include 15 unlensed objects from \citet{ridgway2001} and \citet{kukula2001} and 36 objects (most lensed) from their own observations.  The points in Fig. \ref{fig-peng} are shown with no evolution correction, which explains the (rightward) trend towards more luminous hosts for a given black hole mass as redshift increases.  \citet{peng2006} found that an evolution correction for a passively evolving population formed at high redshift pushes the intermediate-redshift hosts leftward beyond the local relation, implying smaller stellar mass relative to the black hole mass compared to local galaxies.  The constraints from our high-z objects are also dependent on the amount of evolution as discussed above.

\begin{figure}
\epsscale{1.6}
\plotone{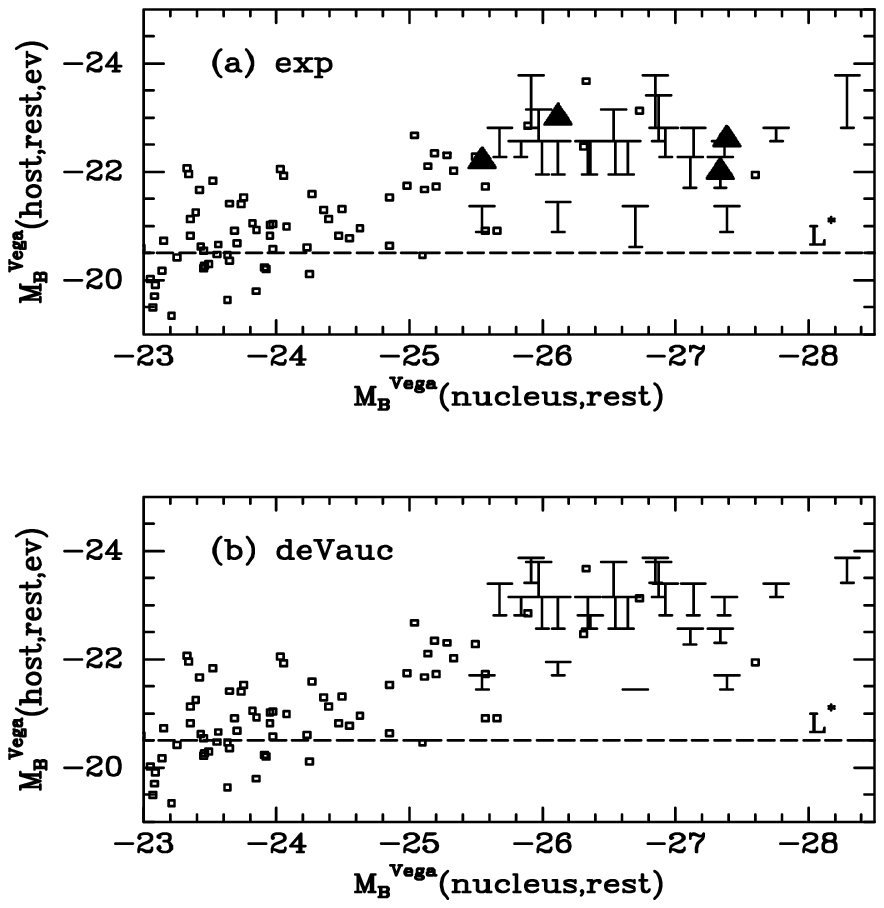}
\caption{Nuclear absolute magnitude v. galaxy magnitude.  Squares show
  local quasars from the compilation of \cite{mm01} adjusted to
  \Hseventy.   For each of our high-z quasars, the two vertical bars
  mark the optimistic (bottom end) and conservative (top end) absolute
  magnitude limits for the range of galaxy scale lengths represented
  in Fig. \ref{fig-isod} and assuming 2 magnitudes of
  evolution. Triangles indicate the four detected hosts at
  z=4. For clarity, the data from ClassicCam are not plotted. \label{fig-allmags}}  
\end{figure}

In Fig. \ref{fig-allmags}, we plot our host limits against rest-frame nuclear $B$-band absolute magnitude, and compare them to those in the $z<0.4$ quasar host galaxy compilation by \cite{mm01}.   
Because the observed $K_s$ directly traces the rest-frame $B$ for our quasars, the $B$-band absolute magnitude plotted here is independent of the nuclear spectral shape
and so provides an complementary approach to using black hole mass as a tracer of the nuclear engine.  
We see from the plots that the limits for exponential hosts imply galaxies fainter than their low-z counterparts, especially using the optimistic limits (bottom bar in each pair) as a bright limit.  The same is true for the optimistic deVaucouleurs limits.  
As in the previous discussion, any excess evolution would push the
two distributions farther apart.  An obvious limitation here is that there are few local quasars with luminosities as high as those of the $z=4$ sample.  
However, one solid conclusion from 
Fig. \ref{fig-allmags} is that there does appear to be a maximum
allowed host.  For the 2 mag of evolution plotted, this maximum
corresponds to $M_B\sim-24$ in the conservative limit, or $M_B\sim-23$
(roughly a 10\lstar\ galaxy) in the optimistic case.  Alternatively,
if there are ever independent suggestions that the mass limit must be
less than that corresponding to an 10\lstar~galaxy, then our results
would imply that the evolution must be more than the 2 mag assumed.

\subsection{K-band Galaxy Evolution:  The K-z relation}

\begin{figure}
\epsscale{0.8}
\plotone{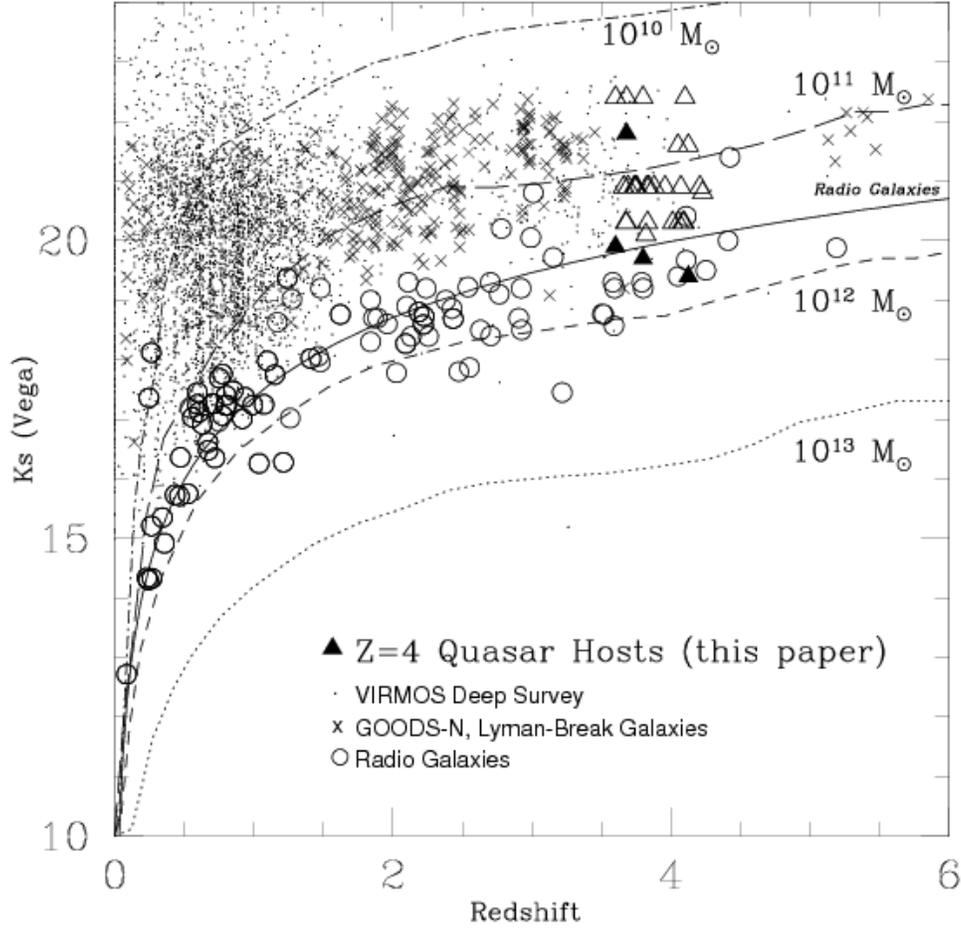}
\caption{K-z Diagram for Radio Galaxies, Field Galaxies, and $z=4$
Quasar Hosts.  The filled triangles are the four detected host
galaxies of this survey;  open triangles are optimistic limits listed 
in Table 4.  Open circles are radio galaxies (Lacy et
al. 2000, De Breuck et al. 2002, Willott et al 2003, De Breuck et al
2006).  X's are optically-selected galaxies (Reddy et al. 2006, McLure
et al 2006), and dots are galaxies in the VIRMOS deep galaxy survey
(Iovino et al. 2005, Temporin et al. 2008). The solid line is the fit
to the locus of radio galaxies, as given by  Willott et al. (2003).
The dotted, dashed, long-dashed and dot-dashed lines are evolutionary
tracks for elliptical galaxies of mass  
10$^{13} M_{\sun}$, 10$^{12} M_{\sun}$,10$^{11} M_{\sun}$, and 
10$^{10} M_{\sun}$, respectively, as computed by Rocca-Volmerange et
al. (2004).  All magnitudes are $K_s$ in the Vega system.  The detected 
quasar hosts have $K_s$ magnitudes consistent with those of
 massive galaxies at their redshifts.
\label{fig-kz}}
\end{figure}

To look at  our observations another way,
we plot $K_s$-magnitude versus redshift in Figure \ref{fig-kz}.  
The observed $K_s$-magnitude of a given galaxy varies with redshift 
because of k-corrections, evolution of the galaxy's stellar population, and
merging.  Here we compare the quasar host galaxies of our study  
with observations of the $K_s$-magnitude of field galaxies at the same redshift.

We plot observed $K_s$-magnitudes for radio galaxies \citep{lacy2000,
debreuck2002, willott2003,debreuck2006}, which define the locus  
of brightest galaxies at all redshifts.  
The locus of radio galaxies is plotted as a solid line given by 
$$ K_s = 17.37 + 4.53 log_{10} z - 0.31 (log_{10}z)^2$$  

\noindent
from \citet{willott2003}.  Fainter galaxies found as Lyman dropouts
\citep{reddy2006} or similar optical selection \citep{iovino2005,  
mclure2006,temporin2008}
and subsequent spectroscopic redshift measurement are shown as well.  
We plot Vega magnitudes for $K_s$, and convert from $K_{AB}$ given in the
literature by assuming that $K_{AB}=K_{Vega}+{1.84}$.  The sharp locus  
of the radio galaxies is interpreted to indicate a maximum mass for
galaxies of $10^{12}\rm M_\sun$, possibly the result of a 
fragmentation limit in cooling proto-galactic gas clouds.  
Also shown are the expected $K_s$-band
evolution tracks for elliptical galaxies of various masses as
computed by \citet{rocca2004}.  The evolution curves for spirals
are similar (see \citet{rocca2004}).
We see that between $z=0.25$ and $z=4$, we expect a given mass galaxy to
undergo approximately 5 magnitudes of
evolution at $K_s$. 

Note that of the four  z=4 hosts detected here, three
(SDSS010905.8+001617.1, SDSSpJ023446.58-001415.9, and PC2047+0123) are
radio quiet.  The other, SDSSpJ084811.52-001418.0, lacks definitive
radio data.  Despite having radio quiet nuclei, two of the 
three most luminous hosts have magnitudes comparable to those of 
radio galaxies at the same redshift.

If we interpret the host galaxy detections with the Rocca-Volmerange
et al. (2004)  model for spheroids in the K-z diagram, then we can
derive the ratio of black hole mass to spheroid mass:  we find a ratio
of 0.016, compared to the local value of $1.4\pm0.4 \times 10^{-3}$
seen locally \citep{har04}.  In other words, given the black hole
masses we inferred from the emission lines, the local relation would
imply more massive host galaxies than we see.   This is what is
predicted by theory \citep{croton2006,somerville2008}.

\subsection{Malmquist Bias in This Sample}\label{sec-malmquist}

\citet{lauer2007} have emphasized that surveys of high redshift
quasars suffer from such strong Malmquist bias that drawing
conclusions about the evolution of host galaxies from small samples
is problematic.  The problem is that we pick targets based on
their quasar luminosity, then look for the host.  Even if there is a
correlation between host galaxy luminosity and black hole luminosity,
the observed host galaxies will be systematically fainter than the
mean relation derived from spheroid velocity dispersions.  This is
because for any plausible galaxy luminosity function, there are many
more faint galaxies than bright ones.  Thus the intrinsic scatter
in the relation causes an excess of low-luminosity galaxies with black
holes big enough to make the sample. 

\begin{figure}
\epsscale{1.2}
\plotone{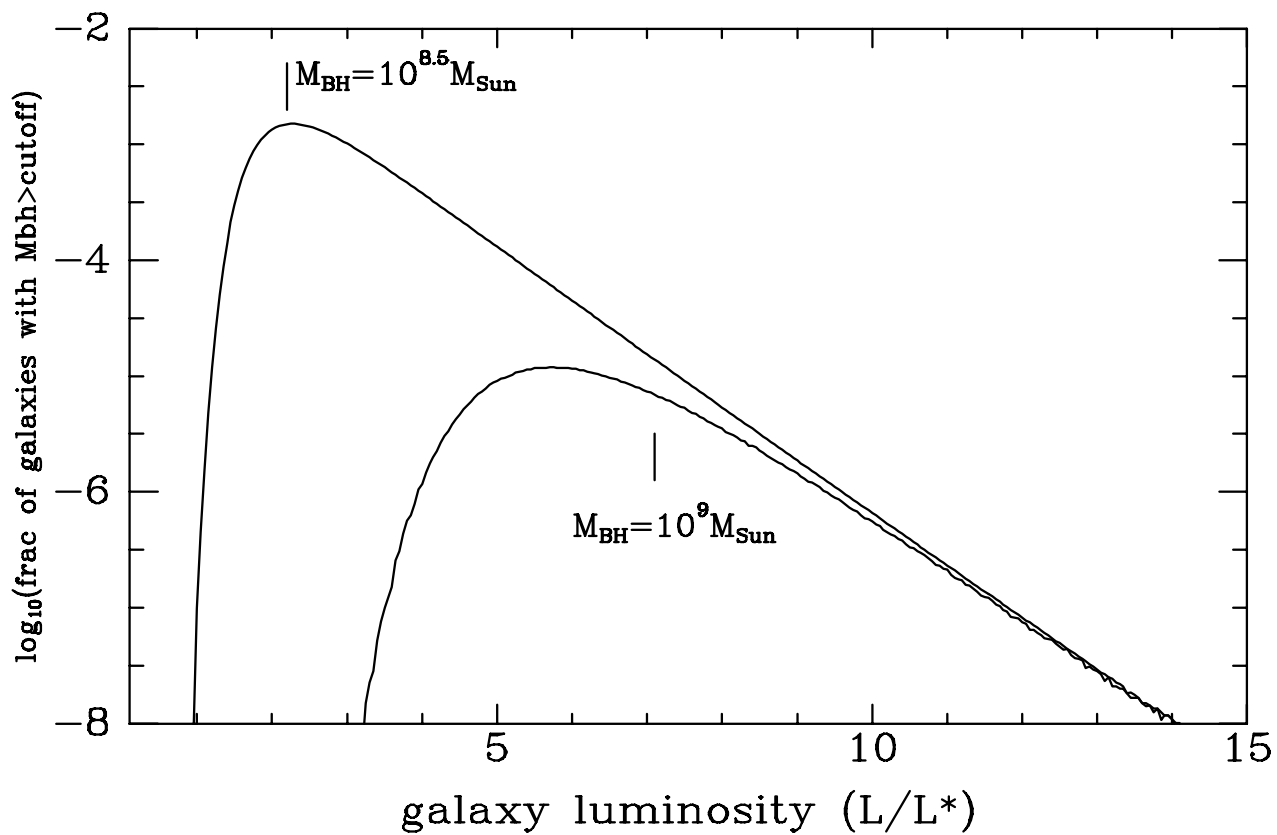}
\caption{Results of Monte Carlo simulation of Malmquist bias for
  M/L=5.  Curves represent the fraction of bright galaxies at each luminosity
  expected to have black hole masses above the cutoff value shown when
  drawn at random from a Schechter function and assigned a black hole
  mass according to the present-day Magorrian relation and its scatter.  The vertical
  bars show the corresponding galaxy mass predicted from the Magorrian
  relation with no scatter.  For the bigger cutoff black hole mass,
  the Malmquist 
  bias is more severe, in that there is a larger contribution from
  galaxies with lower-luminosities than that inferred from Magorrian
  relation mean. 
\label{fig-mult}}
\end{figure}

To look at this effect quantitatively for our sample, we carried out
a Monte-Carlo simulation. 
We chose galaxies randomly from a Schechter galaxy 
luminosity function, from L/ \lstar = 0.1 to 15, given by 

$$\phi(L)dL = {{n_*}(L/\lstar)^\alpha exp(-L/ \lstar)} dL/\lstar$$

\noindent with $\alpha = -0.46$, and $\lstar=2\times10^{10}L_{\odot}$
\citep{sparke2007}.   
We assume a galaxy mass-to-light ratio M/L=5
\citep{cappellari2006,tortora2009}, so that the  corresponding mass of
an \lstar~galaxy is 
M*$\approx10^{11}M_{\odot}$.  Then, we assign to each galaxy a black
hole with mass chosen randomly from the  
\citet{har04} distribution, in which black holes have mass 0.14\%
+/-0.04\% of the galaxy mass.  We further assume that to be found as
a luminous quasar and be eligible for inclusion in our sample, the
object must have a black hole mass above a certain cutoff, taken to
be either $\rm log_{10}(M_{BH}/M_{\odot})= 8.5$ or $9.0$.  
We draw millions of galaxies, and count up the fraction of
host galaxies as a function of host galaxy luminosity for all objects
with black hole masses above our threshold.  The results are shown in
Fig. \ref{fig-mult}.

As expected, we see that for a given value of the black hole cutoff
mass, the results will be skewed towards lower-luminosity hosts than
inferred from the mean value of the black hole-bulge relation.  If
we take a cutoff of $\rm log_{10}(M_{BH}/M_{\odot})= 8.5$, which is at
the small end of our sample, the effect is modest; most of the
hosts would be very close to the expected mean value of about
2\lstar.  However, if we restrict our sample to the more massive black
holes, $\rm log_{10}(M_{BH}/M_{\odot})= 9$, the effect is somewhat larger with
the bulk of the contributors in the range 4-8\lstar, skewed from the
expected mean value of 7\lstar.  However, even if we include the 
effects of the Malquist bias, we could have detected such hosts 
for the most massive black holes in our sample, those with 
$\rm log_{10}(M_{BH}/M_{\odot}) > 9.5$, if the evolution were at least two
magnitudes as discussed above.

\section{Discussion}

When we initiated this program, we hoped to test whether or not
quasars at $z=4$ followed the same correlation of black-hole mass and
host galaxy spheroid mass  seen in spheroids locally.  We knew in
principle that if the very luminous nuclei in the high redshift
objects were hosted by proportionately massive spheroids, they would
be relatively straight-forward to detect at K, with the best
ground-based seeing.  We have taken a very conservative approach to
reducing the data, to looking for host galaxy  detections, and to
estimating host galaxy fluxes and limits.   

We explored more than one way to compare our observed detection of
hosts and limits on host galaxies to the low redshift spheroids.  This
comparison is complicated for a number of reasons, primarily the fact
that we do not measure spheroid velocity dispersion or mass directly,
but must infer the host galaxy properties from the emitted
starlight. Nonetheless, our data indicate that the host galaxies of
some z=4 quasars in our sample are fainter than we expect from the
low-redshift correlations.  

We note that for the mean local values $M_{BH}/M_{gal}=0.14\%$ and
M/L=5, we expect black  holes with $\rm log_{10}(M_{BH}/M_{\odot})= 9.5, 10,
\rm~ and~10.5$, to have host galaxies with 
$M_{gal}=2,7,\rm~and~22x10^{12}\Msun$, or 
$L/\lstar=20,70,\rm~and~220$ respectively.  However, such galaxies are
implausibly large, and do not have present-day analogues.  
For example, as shown in Fig. \ref{fig-kz}, the
upper mass envelope for radio galaxies corresponds to $10^{12}\Msun$,
or $L\approx10\lstar$.  This suggests that the $M_{BH}/M_{gal}$
correlation must plateau at $M_{gal}\approx10^{12}\Msun$.

\section{Summary}

We observed 34 high redshift quasars in the near-IR to
search for their host galaxies.  Our conclusions are the following. 

\noindent
1.  We found that to characterize the PSF and subtract the nuclear
quasar light properly, we had to account for geometric  distortions in
the camera optics, and non-linearity in the detectors,  beyond what is
normally corrected for in standard  pipeline reductions.  We caution
that small uncertainties in the linearity correction, and the use of
PSF stars which are  significantly brighter than the quasar, can lead
to undersubtraction of the nuclear PSF, and spurious host galaxy
detections.  

\noindent
2.  We derived black hole masses for the quasars in the sample from
the profile of the C IV emission line, but noted several cases where
the profile appears to include a narrow and broad component.  Low
signal-to-noise spectra could easily miss the broad  component, with
the result that the black hole mass is underestimated.  The black hole
masses range from $10^{8.7}$ to $10^{10.7}\Msun$.   These quasars are
very luminous, and so rare as to be not represented in some models for
quasar evolution (e.g.  \citet{kh00,dimatteo2008}).  They are more
luminous than the knee in the quasar luminosity function, and  are
probably peak emitters \citep{hopkins2008} that are undergoing a major
merger. 

\noindent
3.  Accretion rates were derived from the observed K photometry, which
directly samples the rest-frame B for the sample quasars.  The median
accretion rate of the sample corresponds to an Eddington fraction of
$L_{bol}/L_{Edd} = 0.41 \pm 0.3$, consistent with the 
findings for other samples such as the Sloan quasars, and for low
redshift quasars whose host galaxies have been studied by  a number of
authors. 

\noindent
4.  We estimated the K-band magnitudes of the host galaxies of the
quasars in our sample with a new method which takes into account  the
isophotal diameter of the galaxies as a function of redshift, as well
as the surface brightness limit  of the images.  We explored parameter
space by considering galaxies with exponential and deVaucouleur radial
light distributions, and quantified the dependance of derived host
galaxy properties on assumed galaxy properties.  

\noindent
5.  We detected host galaxies for 4 quasars, at least three of which are radio
quiet.{\it The
  detections all occurred on our deepest, sharpest images}.  The
K-band luminosities of the hosts are consistent with  massive galaxies
at the redshifts of the quasars.  For one object with H-band data, the
K-H color is bluer than expected for a star-forming galaxy at the
quasar redshift, but the uncertainties in the color are large. 

\noindent
6.  We interpreted the detected hosts and limits on host luminosity in
several ways, taking into account expected evolution of the stellar
populations.  We find that 
if the hosts are already spheroids at early times, then their
black-hole/bulge relation could well be consistent with that for local
galaxies and luminous quasars; our limits are weak in the case of very
compact or very extended spheroids.   On the other hand, if the hosts
are exponential disks, they likely have less stellar mass for a given
black hole mass than would be inferred from the extrapolation of the
local relation to high black hole masses, but they could contain as
much stellar mass as the spheroidal component of local luminous
quasars.  Any $B$-band rest frame luminosity evolution in excess of
the 2 magnitudes assumed would make any evolution in the
black-hole/bulge relation stronger; such would be the case for a
stellar population younger than $\sim$400Myr. 

\noindent
7.  If we interpret the K-magnitudes of the detected hosts with models
for the evolution of massive spheroids we find that the ratio of
black hole mass to spheroid mass for the 4 detected hosts is
approximately 0.02, compared to $1.4\pm0.4 \times 10^{-3}$ observed in
local spheroids.  Several authors have pointed out that the Malmquist
bias  inherent in any study that looks for hosts in very luminous,
rare quasars will overestimate the black hole mass to spheroid  ratio,
given the likely scatter in the correlations, and the fact that faint
galaxies outnumber bright ones.  We made a rough  estimate of the
Malmquist bias in our sample through a Monte Carlo simulation.  We
find that our detection rate is inconsistent with what we should have
seen, had the z=4 quasars followed the same relation between
black-hole mass and  spheroid mass.  Instead, the host galaxies in the
past appear fainter (and by assumption less massive) than  host
galaxies today.  This conclusion depends on uncertain assumptions for
the scatter in the black-hole mass-spheroid  correlation, the
mass-to-light ratios of galaxies, and the evolution of the spectral
energy distribution of spheroids. However, our results are in broad
agreement with semi-analytical models for the growth of black holes
and merger-induced activity.

\acknowledgments
We are grateful to the staffs of the Las Campanas, Gemini,
and Keck Observatories, and to the referee for comments that helped to improve the presentation.
This work was carried out with help from undergraduate students
Francesca D'Arcangelo, Shelby Kimmel, Melissa Rice, Talia Sepersky,
Rebecca Stoll, and Amanda Zangari, who were supported in part by the
Keck Northeast Astronomy Consortium's NSF-REU grant 
AST-0353997.  University of Arizona undergraduates who worked on 
this project were Angela Bivens, who was supported by the NSF program 
Futurebound to Pima Community College,  
and Meri Hidalgo Hembree, who was supported by a UA/NASA Space Grant.
McLeod acknowledges support from the Theresa Mall
Mullarkey Associate Professorship.   
We are grateful to Paul Martini for providing the IRAF {\it panic}
package, 
Brian McLeod for providing {\it imfitfits}, and Ed Olszewski for
carrying out one of the Magellan runs (with thanks to the University
of Michigan Astronomers who gave up some nights). 
The Gemini observations were obtained under program GN-2003B-C-3 (PI:
Bechtold).   This research has made use of SAOImage DS9, developed by
Smithsonian Astrophysical Observatory.  This research has also made extensive
use of the NASA/IPAC Infrared Science Archive, and the NASA/IPAC
Extragalactic Database (NED), both of which are operated by 
the Jet Propulsion Laboratory, California Institute of Technology,
under contract with the National Aeronautics and Space Administration.
The analysis pipeline used to reduce the DEIMOS data was developed at
UC Berkeley with support from
NSF grant AST-0071048.  Thanks to Michael Cooper for help with the
DEIMOS reductions, and Greg Wirth 
and Buell Jannuzi for help with DEIMOS observing.

{\it Facilities:} \facility{Magellan:Baade (PANIC and ClassicCam
near-infrared cameras)}, \facility{Gemini:Gillett (NIRI infrared
camera)}\facility{Keck: (DEIMOS)}

\clearpage

\clearpage

\clearpage

\begin{figure}
\plotone{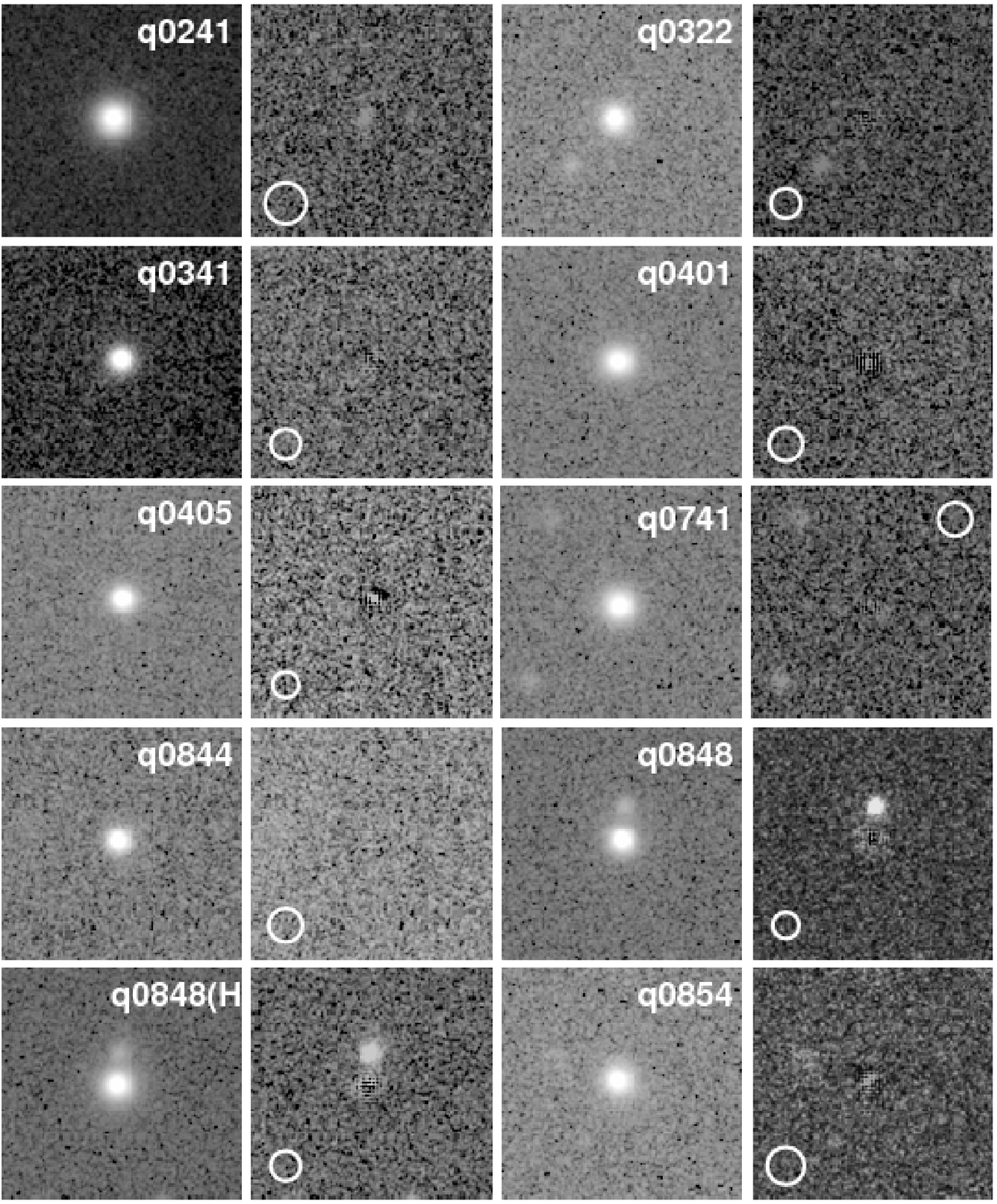}
\noindent Figure \ref{fig-fits} cont'd--targets observed with PANIC.  In most cases the fit to the core is excellent.  
\end{figure}

\begin{figure}
\plotone{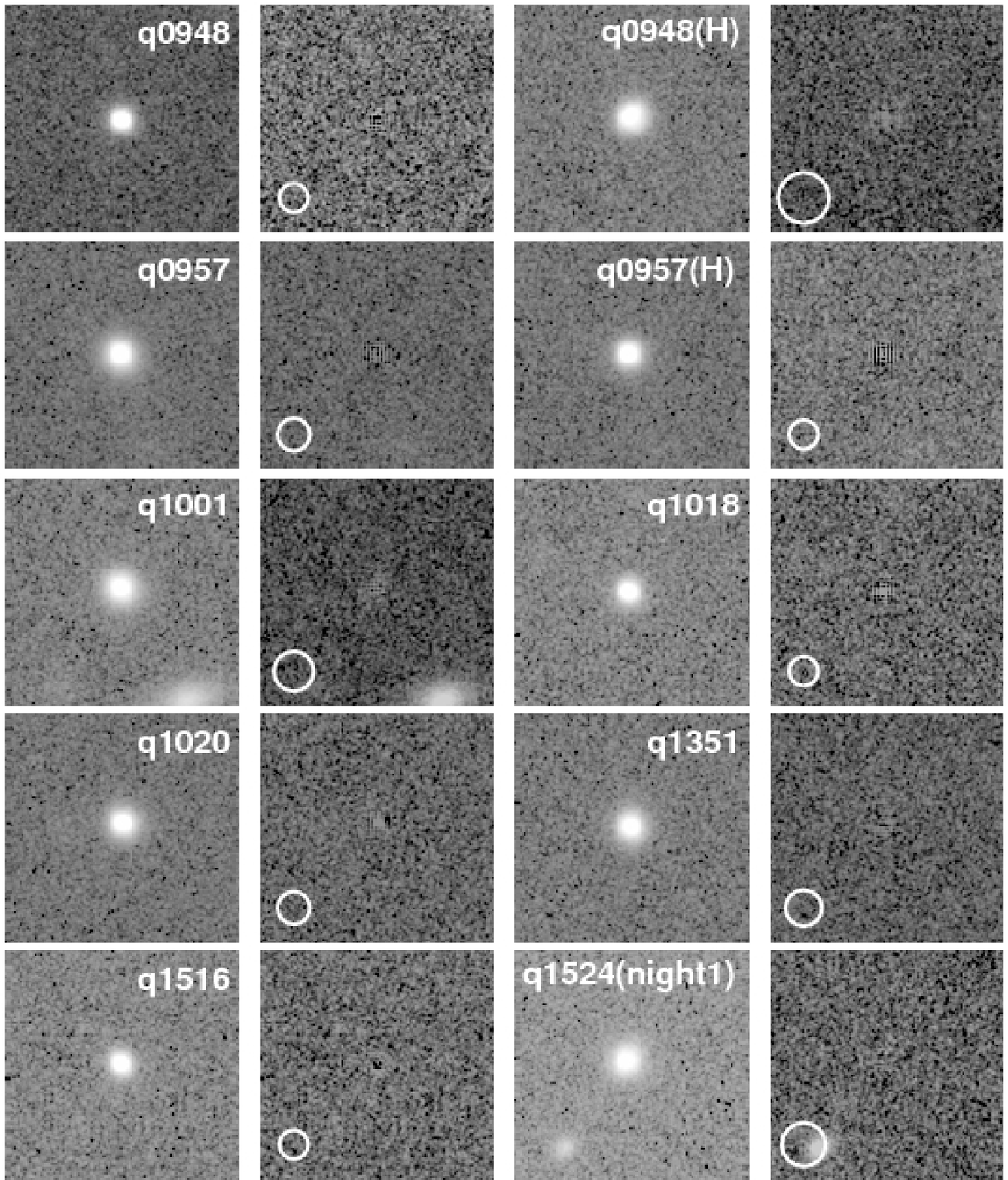}
\noindent Figure \ref{fig-fits} cont'd--targets observed with PANIC.
\end{figure}

\begin{figure}
\plotone{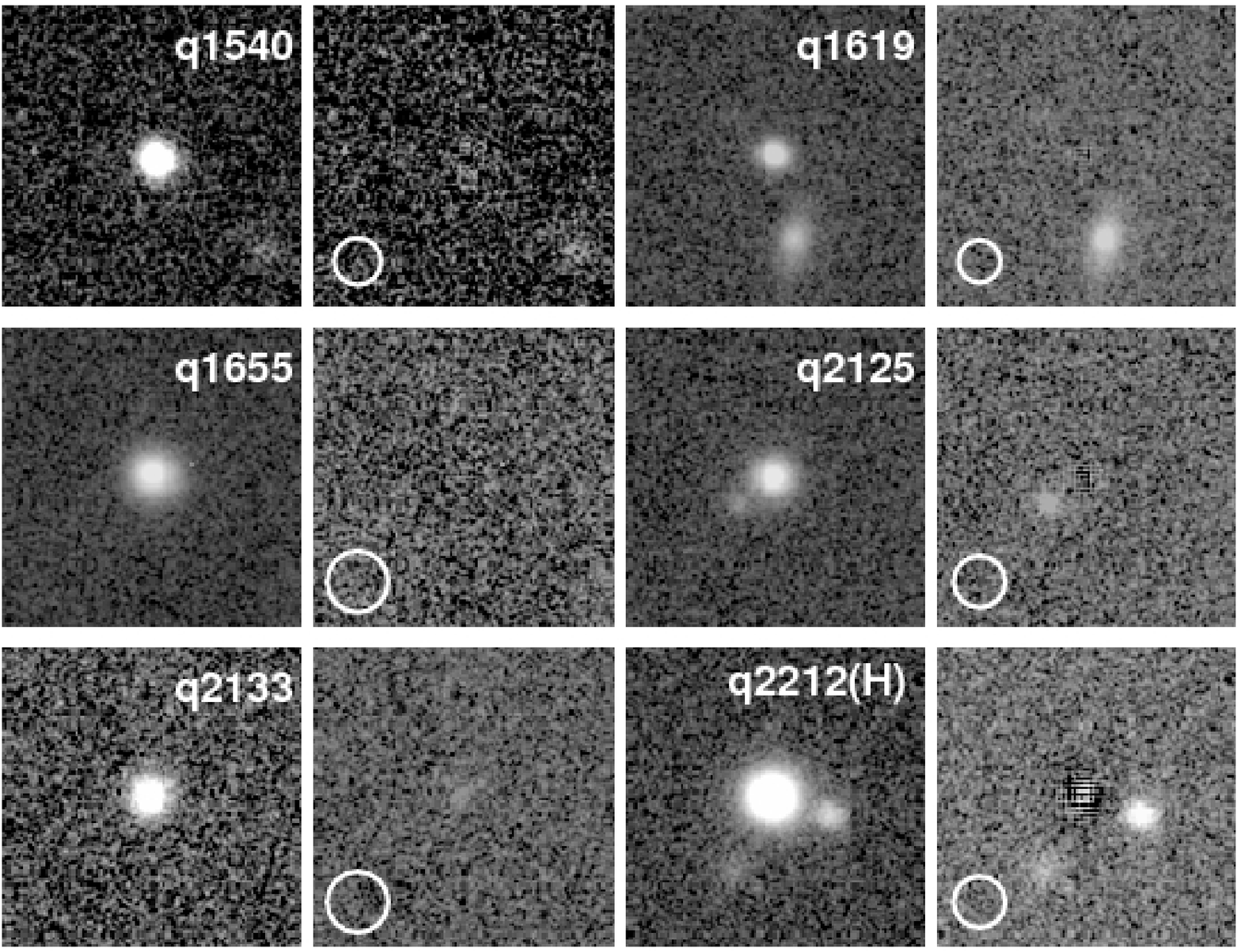}
\noindent Figure \ref{fig-fits} cont'd--targets observed with PANIC.
\end{figure}

\begin{figure}
\plotone{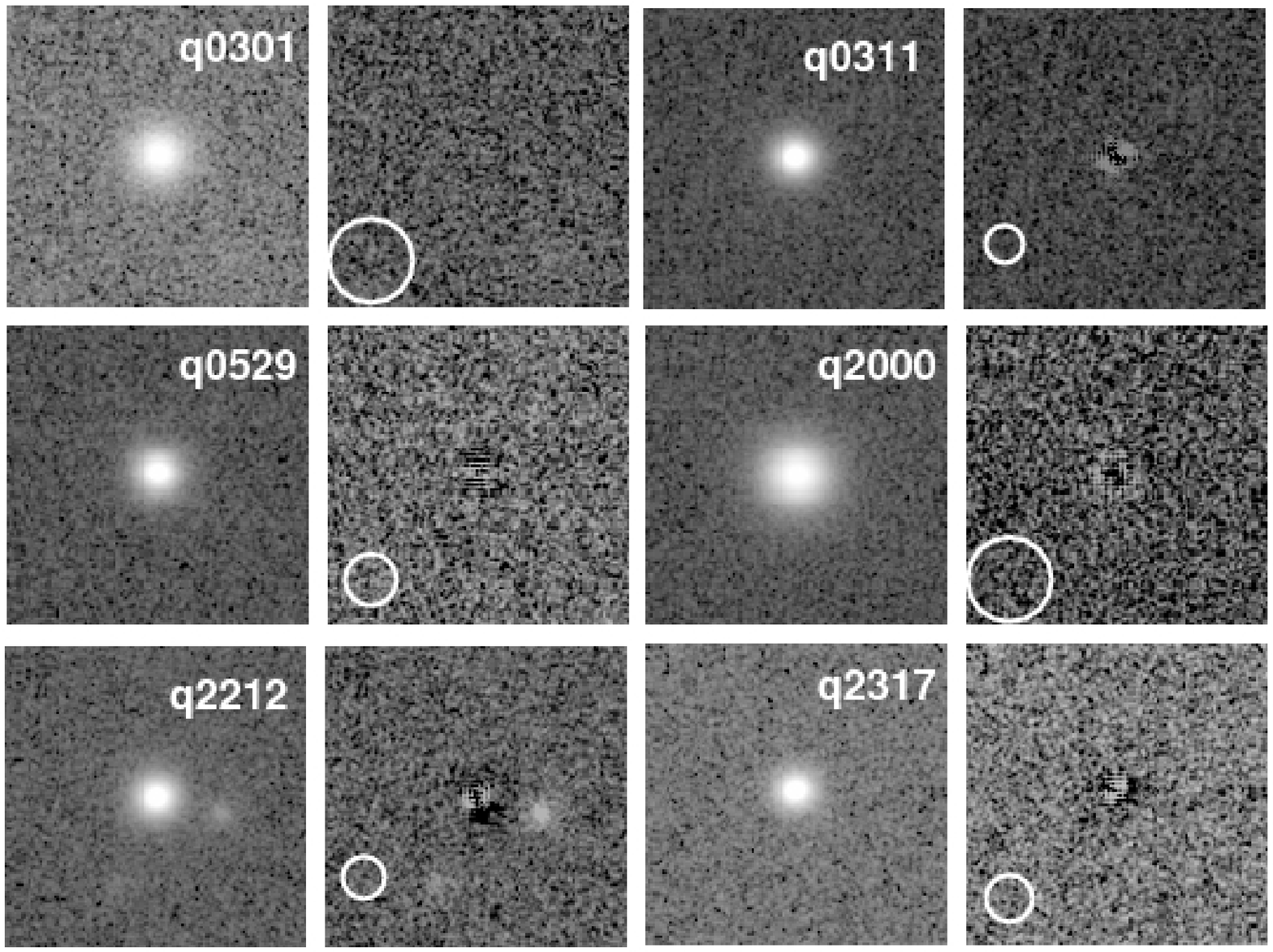}
\noindent Figure \ref{fig-fits} cont'd--targets observed with ClassicCam.  Fits are generally poorer than those for NIRI and PANIC.  As described in the text, for the small ClassicCam field of view, PSF stars were not visible on the same frames as the quasar and had to be obtained in separate observations interleaved with the quasar exposures. In addition, the ClassicCam images generally have broader PSFs and shallower depths.
\end{figure}

\clearpage

\begin{figure}
\includegraphics[scale=0.8,angle=180]{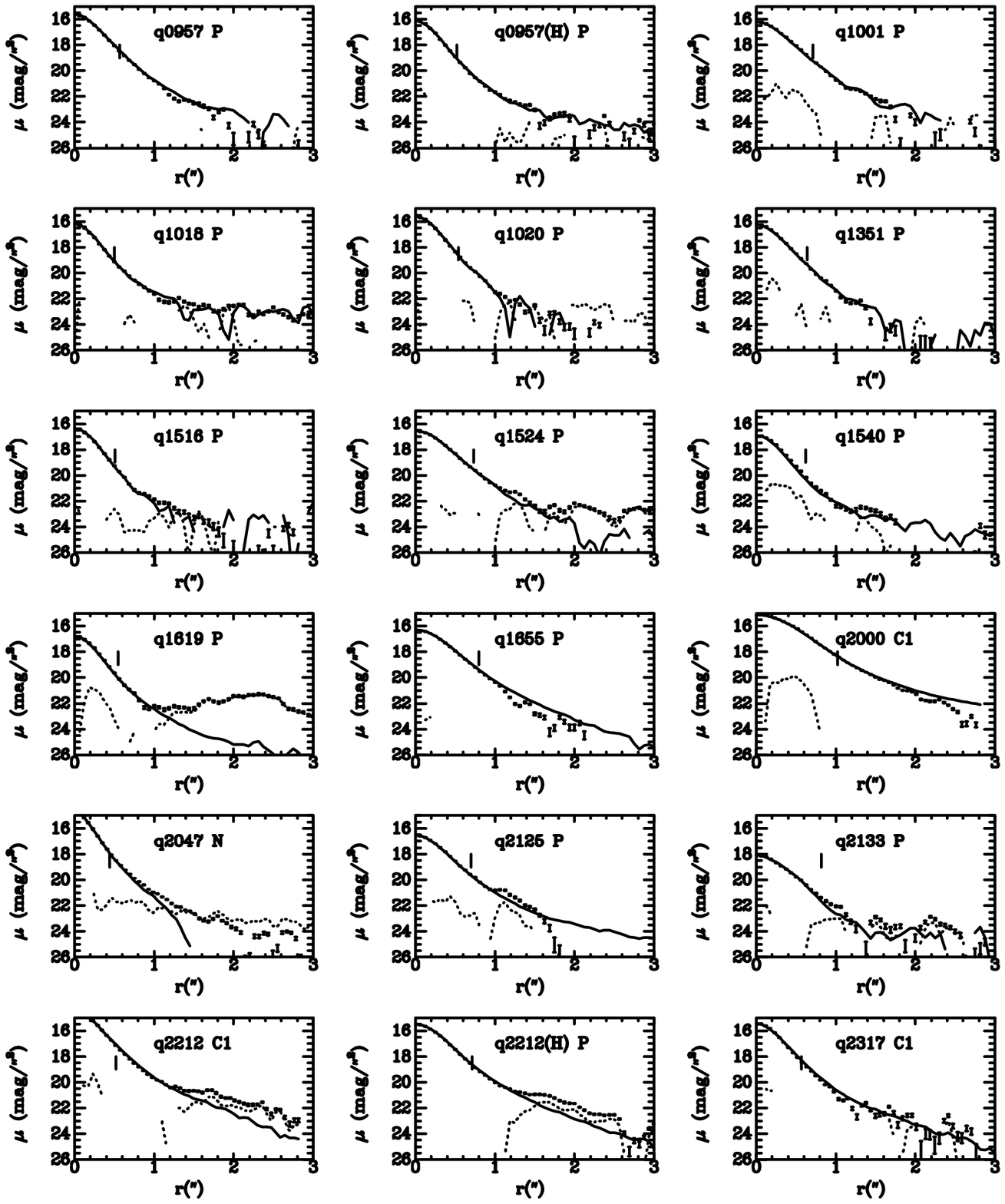}

\noindent Fig. \ref{fig-profiles} cont'd.
\end{figure}



\clearpage
\begin{deluxetable}{llllllllllll}

\tabletypesize{\scriptsize}
\tablecaption{Sample and Observations\label{tab-obs}}
\tablewidth{0pt}

\tablehead{

\colhead{}     & \colhead{}   & \colhead{} & \colhead{} & \colhead{} & \colhead{} & \colhead{} & \colhead{Time} & \colhead{FWHM} & \colhead{SBlim\tablenotemark{c}}\\

\colhead{Name} &  \colhead{z} & \colhead{$\rm r_{AB}$} & \colhead{ref\tablenotemark{a}} & \colhead{Date} & \colhead{Inst\tablenotemark{b}} & \colhead{Filter} & \colhead{(min)} & \colhead{(\arcsec)} & \colhead{$mag/''^2$}

}

\startdata
\cola SDSSpJ010905.81+001617.1\colb 3.680\colc 20.50\cold Fan01,SS\cole 2003 Sep 11\colf N\colg Ks\colh  250\coli  0.35\colj  22.62\eol
\cola SDSSJ012019.99+000735.5\colb 4.101\colc 19.96\cold SS\cole 2003 Nov 05\colf N\colg Ks\colh  125\coli  0.32\colj  22.41\eol
\cola SDSSpJ012650.77+011611.8\colb 3.660\colc 19.20\cold Fan99\cole 2003 Nov 03\colf N\colg Ks\colh  125\coli  0.41\colj  22.36\eol
\cola SDSSpJ023446.58-001415.9\colb 3.600\colc 20.17\cold Fan01,SS\cole 2003 Sep 12\colf N\colg Ks\colh  245\coli  0.39\colj  22.64\eol
\cola BRI0241-0146\colb 4.053\colc 18.50\cold SL96,ACS\cole 2005 Jan 26-27\colf P\colg Ks\colh  143\coli  0.55\colj  21.67\eol
\cola BRI0241-0146\colb 4.053\colc 18.50\cold SL96,ACS\cole 2003 Sep 10\colf N\colg H\colh  219\coli  0.45\colj  23.28\eol
\cola BRJ0301-5537\colb 4.133\colc 19.20\cold P01\cole 2002 Sep 22\colf C1\colg Ks\colh  108\coli  0.82\colj  20.84\eol
\cola BRJ0311-1722\colb 4.039\colc 18.30\cold P01\cole 2002 Sep 21\colf C1\colg Ks\colh  129\coli  0.38\colj  20.83\eol
\cola SDSSpJ031427.92+002339.4\colb 3.680\colc 20.55\cold Fan01,SS\cole 2003 Nov 06\colf N\colg Ks\colh  125\coli  0.57\colj  22.30\eol
\cola SDSSpJ032226.10-055824.7\colb 3.957\colc 20.07\cold SS\cole 2004 Jan 05\colf P\colg Ks\colh  135\coli  0.41\colj  21.76\eol
\cola SDSSpJ034109.35-064805.1\colb 4.070\colc 20.48\cold SS\cole 2004 Jan 04\colf P\colg Ks\colh  135\coli  0.41\colj  21.95\eol
\cola BR0401-1711\colb 4.227\colc 19.60\cold P01\cole 2005 Jan 28\colf P\colg Ks\colh   94\coli  0.43\colj  21.58\eol
\cola SDSSpJ040550.26+005931.2\colb 4.050\colc 20.22\cold Fan01\cole 2004 Jan 02\colf P\colg Ks\colh  128\coli  0.35\colj  21.81\eol
\cola BRJ0529-3552\colb 4.172\colc 18.80\cold P01\cole 2003 Jan 20\colf C2\colg Ks\colh  108\coli  0.42\colj  20.45\eol
\cola PMNJ0741-5305\colb 3.743\colc 21.90\cold L01\cole 2004 Jan 06\colf P\colg Ks\colh  135\coli  0.47\colj  21.87\eol
\cola SDSSpJ084455.08+001848.5\colb 3.690\colc 19.90\cold S01,SS\cole 2004 Jan 05\colf P\colg Ks\colh  142\coli  0.45\colj  21.78\eol
\cola SDSSpJ084811.52-001418.0\colb 4.124\colc 19.24\cold SS\cole 2004 Jan 02\colf P\colg Ks\colh  162\coli  0.36\colj  21.98\eol
\cola SDSSpJ084811.52-001418.0\colb 4.124\colc 19.24\cold SS\cole 2004 Jan 04\colf P\colg Hc\colh  199\coli  0.41\colj  22.82\eol
\cola SDSSpJ085430.18+004213.6\colb 4.084\colc 20.29\cold SS\cole 2004 Jan 03\colf P\colg Ks\colh  155\coli  0.50\colj  21.70\eol
\cola SDSSJ094822.97+005554.3\colb 3.878\colc 20.30\cold SS\cole 2004 Mar 09\colf P\colg Ks\colh  196\coli  0.39\colj  22.00\eol
\cola SDSSJ094822.97+005554.3\colb 3.878\colc 20.30\cold SS\cole 2005 Jan 28\colf P\colg Hc\colh  203\coli  0.70\colj  22.81\eol
\cola SDSSJ095755.63-002027.5\colb 3.753\colc 19.36\cold SS\cole 2005 Jan 26\colf P\colg Ks\colh  250\coli  0.46\colj  22.16\eol
\cola SDSSJ095755.63-002027.5\colb 3.753\colc 19.36\cold SS\cole 2005 Jan 27\colf P\colg Hc\colh  202\coli  0.42\colj  22.32\eol
\cola SDSSJ100151.58-001627.0\colb 3.674\colc 19.31\cold SS\cole 2004 Mar 11\colf P\colg Ks\colh  169\coli  0.57\colj  21.85\eol
\cola SDSSpJ101832.46+001436.4\colb 3.826\colc 20.22\cold SS\cole 2004 Mar 10\colf P\colg Ks\colh  209\coli  0.40\colj  21.87\eol
\cola SDSSpJ102043.82+000105.8\colb 4.207\colc 20.53\cold SS\cole 2004 Mar 08\colf P\colg Ks\colh  182\coli  0.43\colj  21.88\eol
\cola SDSSJ135134.46-003652.2\colb 4.010\colc 20.13\cold SS\cole 2004 Mar 11\colf P\colg Ks\colh  209\coli  0.51\colj  21.92\eol
\cola SDSSJ151618.44-000544.1\colb 3.735\colc 20.23\cold SS\cole 2004 Mar 08\colf P\colg Ks\colh  148\coli  0.41\colj  21.86\eol
\cola SDSSpJ152443.19+011358.9\colb 4.095\colc 20.14\cold SDSS2\cole 2003 Jul 17-18\colf P\colg Ks\colh  213\coli  0.58\colj  21.67\eol
\cola SDSSpJ154014.57+001854.7\colb 3.830\colc 20.41\cold S01,SS\cole 2004 Mar 10\colf P\colg Ks\colh  169\coli  0.50\colj  21.89\eol
\cola SDSSpJ161926.87-011825.2\colb 3.840\colc 20.17\cold Fan00\cole 2004 Mar 09\colf P\colg Ks\colh  155\coli  0.44\colj  21.90\eol
\cola SDSSpJ165527.61-000619.2\colb 3.990\colc 20.05\cold Fan00\cole 2003 Jul 19\colf P\colg Ks\colh  157\coli  0.64\colj  21.67\eol
\cola [HB89]2000-330\colb 3.773\colc 17.6\cold HB,WFPC2\cole 2002 Sep 22\colf C1\colg Ks\colh  117\coli  0.82\colj  20.72\eol
\cola PC2047+0123\colb 3.799\colc 19.46\cold S91\cole 2003 Sep 10-11\colf N\colg Ks\colh  154\coli  0.35\colj  22.46\eol
\cola [VH95]2125-4529\colb 3.763\tablenotemark{d}\colc 19.50\cold HV96\cole 2003 Jul 18\colf P\colg Ks\colh  146\coli  0.56\colj  21.54\eol
\cola [VCV96]Q2133-4625\colb 4.180\colc 21.23\cold VV,HV96\cole 2003 Jul 17\colf P\colg Ks\colh  243\coli  0.66\colj  21.84\eol
\cola BR2212-1626\colb 3.990\colc 18.80\cold SL96\cole 2002 Sep 21\colf C1\colg Ks\colh  117\coli  0.42\colj  20.69\eol
\cola BR2212-1626\colb 3.990\colc 18.80\cold SL96\cole 2003 Jul 19\colf P\colg Hc\colh  157\coli  0.57\colj  21.89\eol
\cola BRJ2317-4345\colb 3.943\colc 18.80\cold P01\cole 2002 Sep 20\colf C1\colg Ks\colh  117\coli  0.46\colj  20.69\eol

\enddata
\tablenotetext{a}{Sources for $z$,$r_{AB}$: SS (SDSS DR6); Fan99,00,01 \citep{fan99,fan00,fan01}; L01 \citep{landt01}; P01 \citep{peroux01}; S01 \citep{schneider01}; SL96 \citep{sl96}; HB \citep{hb89}; VV \citep{vv96,vv01};  ACS,WFPC2 archival HST images measured here; HV96 \citep{hv96} and $r_{AB} = R + 0.3$.}
\tablenotetext{b}{C1,C2=ClassicCam with scale 0\farcs115/pix, 0\farcs95/pix; N=NIRI with scale 0\farcs116/pix; P=PANIC with scale 0\farcs125/pix}
\tablenotetext{c}{Surface brightness limit of final image (Vega mags) derived from 1$\sigma$ pixel-to-pixel variation.  Effective limits for the magnified images used with PSF fits are $\sim0.4$mag brighter.}
\tablenotetext{d}{Improved redshift from Keck spectrum in this paper}
\end{deluxetable}

\begin{deluxetable}{lrrrllrlrrl}

\tabletypesize{\scriptsize}
\tablecaption{Radio Properties\label{tab-radio}}
\tablewidth{0pt}

\tablehead{

\colhead{} &  \colhead{} & \colhead{} & \colhead{} & \colhead{} & \colhead{$F_{1.4GHz}$\tablenotemark{c}} & \colhead{} & \colhead{} &\colhead{} & \colhead {RL or}\\

\colhead{Name} &  \colhead{z} & \colhead{E(B-V)} & \colhead{$i^{AB}$\tablenotemark{a}} & \colhead{i ref\tablenotemark{b}} & \colhead{(mJy)} & \colhead{Flux ref\tablenotemark{d}} & \colhead{t} &\colhead{R $(2\sigma)$\tablenotemark{e}} & \colhead {RQ?}

}

\startdata
\cola SDSSpJ010905.81+001617.1\colb 3.680\colc 0.026\cold 20.31\cole SS\colf $<$ 0.110\colh FIRST\coli 18.80\colj  0.90\colk Q\eol
\cola SDSSJ012019.99+000735.5\colb 4.101\colc 0.037\cold 19.82\cole SS\colf $<$ 0.112\colh FIRST\coli 18.78\colj  0.72\colk Q\eol
\cola SDSSpJ012650.77+011611.8\colb 3.660\colc 0.026\cold 19.17\cole Fan99\colf 0.210\colh C01\coli 18.09\colj  0.43\colk Q\eol
\cola SDSSpJ023446.58-001415.9\colb 3.600\colc 0.022\cold 19.88\cole SS\colf $<$ 0.108\colh FIRST\coli 18.82\colj  0.73\colk Q\eol
\cola BRI0241-0146\colb 4.053\colc 0.028\cold 18.24\cole ACS\colf $<$ 0.150\colh FIRST\coli 18.46\colj  0.13\colk Q\eol
\cola BRJ0301-5537\colb 4.133\colc 0.010\cold 18.97\cole P01\colf ...\colh [NED]\coli ...\colj ...\colk ?\eol
\cola BRJ0311-1722\colb 4.039\colc 0.032\cold 18.01\cole P01\colf $<$ 2.500\colh NVSS\coli 15.41\colj  1.34\colk ?\eol
\cola SDSSpJ031427.92+002339.4\colb 3.680\colc 0.100\cold 20.04\cole SS\colf $<$ 0.112\colh FIRST\coli 18.78\colj  0.81\colk Q\eol
\cola SDSSpJ032226.10-055824.7\colb 3.957\colc 0.053\cold 19.86\cole SS\colf $<$ 2.500\colh NVSS\coli 15.41\colj  2.08\colk ?\eol
\cola SDSSpJ034109.35-064805.1\colb 4.070\colc 0.053\cold 20.09\cole SS\colf $<$ 2.500\colh NVSS\coli 15.41\colj  2.17\colk ?\eol
\cola BR0401-1711\colb 4.227\colc 0.026\cold 19.33\cole P01\colf $<$ 2.500\colh NVSS\coli 15.41\colj  1.87\colk ?\eol
\cola SDSSpJ040550.26+005931.2\colb 4.050\colc 0.444\cold 19.08\cole Fan01\colf $<$ 2.500\colh NVSS\coli 15.41\colj  1.77\colk ?\eol
\cola BRJ0529-3552\colb 4.172\colc 0.031\cold 18.51\cole P01\colf $<$ 2.500\colh NVSS\coli 15.41\colj  1.54\colk ?\eol
\cola PMNJ0741-5305\colb 3.743\colc 0.227\cold 21.08\cole L01\colf 44.000\colh G93\coli 12.29\colj  3.51\colk L\eol
\cola SDSSpJ084455.08+001848.5\colb 3.690\colc 0.035\cold 19.74\cole SS\colf 5.830\colh NVSS\coli 14.49\colj  2.10\colk L\eol
\cola SDSSpJ084811.52-001418.0\colb 4.124\colc 0.030\cold 18.93\cole SS\colf $<$ 2.500\colh NVSS\coli 15.41\colj  1.71\colk ?\eol
\cola SDSSpJ085430.18+004213.6\colb 4.084\colc 0.049\cold 19.84\cole SS\colf $<$ 2.500\colh NVSS\coli 15.41\colj  2.07\colk ?\eol
\cola SDSSJ094822.97+005554.3\colb 3.878\colc 0.109\cold 19.85\cole SS\colf 2.320\colh FIRST\coli 15.49\colj  1.75\colk L\eol
\cola SDSSJ095755.63-002027.5\colb 3.753\colc 0.033\cold 19.28\cole SS\colf $<$ 0.149\colh FIRST\coli 18.47\colj  0.63\colk Q\eol
\cola SDSSJ100151.58-001627.0\colb 3.674\colc 0.037\cold 19.18\cole SS\colf $<$ 0.146\colh FIRST\coli 18.49\colj  0.58\colk Q\eol
\cola SDSSpJ101832.46+001436.4\colb 3.826\colc 0.043\cold 19.98\cole SS\colf $<$ 0.151\colh FIRST\coli 18.45\colj  0.91\colk Q\eol
\cola SDSSpJ102043.82+000105.8\colb 4.207\colc 0.040\cold 19.75\cole SS\colf 2.000\colh FIRSTc\coli 15.65\colj  1.64\colk L\eol
\cola SDSSJ135134.46-003652.2\colb 4.010\colc 0.034\cold 19.79\cole SS\colf $<$ 0.148\colh FIRST\coli 18.47\colj  0.83\colk Q\eol
\cola SDSSJ151618.44-000544.1\colb 3.735\colc 0.057\cold 19.89\cole SS\colf $<$ 0.045\colh C01\coli 19.77\colj  0.35\colk Q\eol
\cola SDSSpJ152443.19+011358.9\colb 4.095\colc 0.055\cold 19.88\cole SS\colf $<$ 0.143\colh FIRST\coli 18.51\colj  0.85\colk Q\eol
\cola SDSSpJ154014.57+001854.7\colb 3.830\colc 0.092\cold 20.06\cole SS\colf $<$ 0.151\colh FIRST\coli 18.45\colj  0.94\colk Q\eol
\cola SDSSpJ161926.87-011825.2\colb 3.840\colc 0.117\cold 19.69\cole Fan00\colf $<$ 0.025\colh C01\coli 20.41\colj  0.01\colk Q\eol
\cola SDSSpJ165527.61-000619.2\colb 3.990\colc 0.298\cold 19.54\cole Fan00\colf 0.010\colh C01\coli 21.40\colj -0.74\colk Q\eol
\cola [HB89]2000-330\colb 3.773\colc 0.130\cold 17.04\cole WFPC2\colf 570.00\colh W90\coli  9.51\colj  3.01\colk L\eol
\cola PC2047+0123\colb 3.799\colc 0.105\cold 19.07\cole S91\colf 0.105\colh S92\coli 18.85\colj  0.09\colk Q\eol
\cola [VH95]2125-4529\colb 3.763\colc 0.022\cold 19.15\cole HV96\colf ...\colh [NED]\coli ...\colj ...\colk ?\eol
\cola [VCV96]Q2133-4625\colb 4.180\colc 0.025\cold 21.04\cole HV96\colf ...\colh [NED]\coli ...\colj ...\colk ?\eol
\cola BR2212-1626\colb 3.990\colc 0.028\cold 18.49\cole SL96\colf $<$ 2.500\colh NVSS\coli 15.41\colj  1.54\colk ?\eol
\cola BRJ2317-4345\colb 3.943\colc 0.011\cold 18.57\cole P01\colf ...\colh [NED]\coli ...\colj ...\colk ?\eol

\enddata
\tablenotetext{a}{Corrected for reddening}
\tablenotetext{b}{SS (SDSS DR6); Fan99,00,01 \citep{fan99,fan00,fan01}; L01 \citep{landt01} computed from $B_j$; P01 \citep{peroux01} computed from $R$; S91 \citep{schneider91} computed from $AB1450$; SL96 \citep{sl96} computed from $AB7000\AA$; HV96 \citep{hv96}; ACS,WFPC archival image measured here}
\tablenotetext{c}{Detections or $1\sigma$ upper limits}
\tablenotetext{d}{FIRST and NVSS were accessed through NED; C01 \citep{carilli01}; G93 \citep{gw93}; S92 \citep{schneider92}; W90 \citep{wo90}}
\tablenotetext{e}{Radio-to-optical flux ratio calculated from detections, or from $2\sigma$ limits for nondetections}

\end{deluxetable}

\begin{deluxetable}{lrrlllrrrr}

\tabletypesize{\scriptsize}
\tablecaption{Black Hole Properties\label{tab-blackholes}}
\tablewidth{0pt}

\tablehead{

\colhead{} &  \colhead{} & \colhead{} & 
\colhead{} & \colhead{$\rm FWHM_{CIV}$\tablenotemark{c}} & 
\colhead{} & \colhead{} & 
\colhead{$\rm M_{BH}$\tablenotemark{e}} & \colhead{$L_{bol}/L_{Edd}$\tablenotemark{e}} & 
\colhead{$\rm M_{BH}$\tablenotemark{f}}\\

\colhead{Name} &  \colhead{z} & \colhead{$\rm AB_{1450}$\tablenotemark{a}} &
\colhead{AB ref}\tablenotemark{b} & \colhead{($km~s^{-1}$)} &
\colhead{CIV ref}\tablenotemark{b} & \colhead{$M_B^{Vega}$\tablenotemark{d}}& 
\colhead{CIV} & \colhead{CIV} &
\colhead{$0.4L_{Edd}$}

}

\startdata
\cola SDSSpJ010905.81+001617.1\colb 3.680\colc 20.89\cold Fan01\cole  4980\colf Fan01\colg  -25.5\colh  9.1\coli 0.40\colj  9.0\eol
\cola SDSSJ012019.99+000735.5\colb 4.101\colc 19.96\cold Fan01\cole  4840\colf SS\colg  -26.7\colh  9.3\coli 0.69\colj  9.4\eol
\cola SDSSpJ012650.77+011611.8\colb 3.660\colc 19.58\cold Fan99\cole  3991/8756\colf Fan99\colg  -26.6\colh  9.2/9.8\coli 0.89/0.19\colj  9.4\eol
\cola SDSSpJ023446.58-001415.9\colb 3.600\colc 20.40\cold S01\cole  3030\colf SS\colg  -26.1\colh  8.7\coli 1.45\colj  9.2\eol
\cola BRI0241-0146\colb 4.053\colc 18.23\cold i\cole  9552\colf C02\colg  -28.3\colh 10.2\coli 0.34\colj 10.1\eol
\cola BRJ0301-5537\colb 4.133\colc 18.97\cold i\cole  6728\colf P01\colg  -27.3\colh  9.8\coli 0.39\colj  9.7\eol
\cola BRJ0311-1722\colb 4.039\colc 18.00\cold i\cole 15117\colf P01\colg  -28.5\colh 10.7\coli 0.15\colj 10.2\eol
\cola SDSSpJ031427.92+002339.4\colb 3.680\colc 20.33\cold Fan01\cole  3758\colf SS\colg  -25.9\colh  8.9\coli 0.74\colj  9.1\eol
\cola SDSSpJ032226.10-055824.7\colb 3.957\colc 20.21\cold SS\cole  7100\colf SS\colg  -26.5\colh  9.6\coli 0.33\colj  9.4\eol
\cola SDSSpJ034109.35-064805.1\colb 4.070\colc 20.70\cold SS\cole  7033\colf SS\colg  -26.0\colh  9.5\coli 0.25\colj  9.2\eol
\cola BR0401-1711\colb 4.227\colc 19.34\cold i\cole  2004/5504\tablenotemark{g}\colf C02\colg  -27.8\colh  8.7/9.5\coli 7.83/1.04\colj  9.9\eol
\cola SDSSpJ040550.26+005931.2\colb 4.050\colc 19.27\cold Fan01\cole  3070\colf Fan01\colg  -27.1\colh  9.0\coli 1.83\colj  9.6\eol
\cola BRJ0529-3552\colb 4.172\colc 18.52\cold i\cole  3210\tablenotemark{g}\colf P01\colg  -27.9\colh  9.2\coli 2.40\colj  9.9\eol
\cola PMNJ0741-5305\colb 3.743\colc 21.04\cold i\cole  9990\colf L01\colg  -26.9\colh  9.7\coli 0.37\colj  9.5\eol
\cola SDSSpJ084455.08+001848.5\colb 3.690\colc 20.05\cold S01\cole  1752/8605\colf Keck\colg  -25.7\colh  8.3/9.7\coli 2.4/0.1\colj  9.0\eol
\cola SDSSpJ084811.52-001418.0\colb 4.124\colc 18.93\cold SS\cole  5371\colf Keck\colg  -27.3\colh  9.6\coli 0.62\colj  9.7\eol
\cola SDSSpJ085430.18+004213.6\colb 4.084\colc 20.10\cold SS\cole  5580\colf Keck\colg  -26.5\colh  9.4\coli 0.49\colj  9.4\eol
\cola SDSSJ094822.97+005554.3\colb 3.878\colc 21.05\cold SS\cole  6078\colf SS\colg  -26.4\colh  9.2\coli 0.58\colj  9.3\eol
\cola SDSSJ095755.63-002027.5\colb 3.753\colc 20.10\cold SS\cole  4964\colf SS\colg  -27.1\colh  9.2\coli 1.15\colj  9.6\eol
\cola SDSSJ100151.58-001627.0\colb 3.674\colc 20.08\cold SS\cole  6813\colf SS\colg  -26.8\colh  9.5\coli 0.47\colj  9.5\eol
\cola SDSSpJ101832.46+001436.4\colb 3.826\colc 20.19\cold S01\cole  5657\colf SS\colg  -26.3\colh  9.3\coli 0.44\colj  9.3\eol
\cola SDSSpJ102043.82+000105.8\colb 4.207\colc 20.35\cold SS\cole  8847\colf SS\colg  -27.4\colh  9.7\coli 0.46\colj  9.7\eol
\cola SDSSJ135134.46-003652.2\colb 4.010\colc 20.74\cold SS\cole  5420\colf SS\colg  -26.9\colh  9.2\coli 0.98\colj  9.5\eol
\cola SDSSJ151618.44-000544.1\colb 3.735\colc 20.10\cold Fan00\cole 10631\colf SS\colg  -26.1\colh  9.9\coli 0.10\colj  9.2\eol
\cola SDSSpJ152443.19+011358.9\colb 4.095\colc 20.73\cold SS\cole 10228\colf SS\colg  -26.9\colh  9.8\coli 0.28\colj  9.5\eol
\cola SDSSpJ154014.57+001854.7\colb 3.830\colc 20.20\cold S01\cole  9707\colf S01\colg  -26.0\colh  9.8\coli 0.11\colj  9.1\eol
\cola SDSSpJ161926.87-011825.2\colb 3.840\colc 20.04\cold Fan00\cole  6076\colf Fan00\colg  -25.8\colh  9.4\coli 0.22\colj  9.1\eol
\cola SDSSpJ165527.61-000619.2\colb 3.990\colc 20.22\cold Fan00\cole  3102/12181\tablenotemark{g}\colf Fan00\colg  -27.2\colh  8.8/10.0\coli 3.21/0.21\colj  9.6\eol
\cola [HB89]2000-330\colb 3.773\colc 17.01\cold i\cole  4081\colf K01\colg  -28.7\colh  9.7\coli 1.55\colj 10.2\eol
\cola PC2047+0123\colb 3.799\colc 19.23\cold S92\cole  6316\colf Keck\colg  -27.4\colh  9.6\coli 0.58\colj  9.7\eol
\cola [VH95]2125-4529\colb 3.763\colc 19.12\cold i\cole 6257\colf Keck\colg   -26.6\colh  9.7\coli 0.28\colj  9.4\eol
\cola [VCV96]Q2133-4625\colb 4.180\colc 21.04\cold D03\cole ...\colf ...\colg  -25.5\colh ...\coli ...\colj  8.9\eol
\cola BR2212-1626\colb 3.990\colc 18.52\cold SL96\cole  2739/6185\tablenotemark{g}\colf C02\colg   -28.5\colh   9.1/9.8\coli 5.59/1.10\colj 10.1\eol
\cola BRJ2317-4345\colb 3.943\colc 18.55\cold i\cole  4612\colf P01\colg  -27.4\colh  9.5\coli 0.79\colj  9.7\eol

\enddata
\tablenotetext{a}{Reddening-corrected AB magnitude at $1450\AA(1+z)$}
\tablenotetext{b}{Sources for spectra and photometry: i (derived from $i^{AB}$ in Table \ref{tab-radio}); C02 \citep{constantin2002}; D03 \citep{dietrich2003}; Fan99,00,01 \citep{fan99,fan00,fan01}; K01 \citep{kuhn2001}; L01 \citep{landt01}; P01 \citep{peroux01}; S01,92 \citep{schneider01,schneider92}; SL96 \citep{sl96}; SS (SDSS DR6); SW00 \citep{sw2000}; Keck this paper}
\tablenotetext{c}{FWHM measured by us from spectra. For objects with two values listed we measured both a broad and
a narrow component.}  
\tablenotetext{d}{Quasar absolute magnitude $M_B$ (Vega system) computed from the observed K-band magnitudes in Table \ref{tab-hosts}}
\tablenotetext{e}{Black hole mass $log_{10}(\rm M_{BH}/M_\odot)$ (derived from $\rm FHWM_{CIV}$  and $\rm AB_{1450}$)  and Eddington fraction $L_{bol}/L_{Edd}$ (from mass and $M_B$).}
\tablenotetext{f}{$log_{10}(\rm M_{BH}/M_\odot)$ calculated from quasar magnitude $M_B$ assuming $L/L_{Edd}=0.4$}
\tablenotetext{g}{Notes: q0405 and q1655 profiles have broad wings; q0529 spectrum is noisy; q2212 is gravitationally lensed \citep{warren2001}}
\end{deluxetable}

\begin{deluxetable}{lrcrrrrrccrlr}

\tabletypesize{\scriptsize}
\tablecaption{Results from $K_s$ Imaging\label{tab-hosts}}
\tablewidth{0pt}

\tablehead{

\colhead{} &  \colhead{} & \colhead{}&
\multicolumn{2}{c}{$L_{exp}/\lstar$\tablenotemark{a}} &
\multicolumn{2}{c}{$L_{deV}/\lstar$\tablenotemark{b}} & \colhead{} &
\colhead{} & \colhead{} & \multicolumn{2}{c}{$L_{host}/\lstar$\tablenotemark{d}} & \colhead{} \\

\colhead{Name} & \colhead{z} & \colhead{$K_{s,quasar}^{Vega}$} &
\colhead{min} & \colhead{max} & 
\colhead{min} & \colhead{max} & 
\colhead{$K_{s,min}^{Vega}$\tablenotemark{c}} &\colhead{Host} & 
\colhead{$K_{s,host}^{Vega}$} &
\colhead{no ev} & \colhead {ev} & \colhead{$M_{B,host}^{Vega}$\tablenotemark{e}}
}
\startdata

\cola SDSSpJ010905.81+001617.1\colb 3.680\colc 18.3\cold    9\cole   14\colf   15\colg   19\colh 21.40\coli y\colj 21.8 &30 &5 &-22.2 \eol
\cola SDSSJ012019.99+000735.5\colb 4.101\colc 17.5\cold    7\cole   14\colf   15\colg   15\colh 21.80\coli n\colj ...&...&...&... \eol
\cola SDSSpJ012650.77+011611.8\colb 3.660\colc 17.1\cold   24\cole   42\colf   42\colg   72\colh 19.90\coli  ?n\colj ...&...&...&... \eol
\cola SDSSpJ023446.58-001415.9\colb 3.600\colc 17.6\cold    9\cole   15\colf   19\colg   24\colh 20.80\coli y\colj 19.9&60 & 10 & -23.0\eol
\cola BRI0241-0146\colb 4.053\colc 15.9\cold   53\cole  130\colf   92\colg  140\colh 18.40\coli   n\colj ...&...&...&... \eol
\cola BRI0241-0146\colb 4.053\colc 16.5\cold ...\cole ...\colf ...\colg ...\colh ...\coli   n\colj ...&...&...&... \eol
\cola BRJ0301-5537\colb 4.133\colc 16.9\cold ...\cole ...\colf ...\colg ...\colh ...\coli   n\colj ...&...&...&... \eol
\cola BRJ0311-1722\colb 4.039\colc 15.6\cold ...\cole ...\colf ...\colg ...\colh ...\coli   n\colj ...&...&...&... \eol
\cola SDSSpJ031427.92+002339.4\colb 3.680\colc 17.9\cold   53\cole  130\colf   92\colg  140\colh 18.40\coli   n\colj ...&...&...&... \eol
\cola SDSSpJ032226.10-055824.7\colb 3.957\colc 17.5\cold   24\cole   42\colf   42\colg   72\colh 19.90\coli   n\colj ...&...&...&... \eol
\cola SDSSpJ034109.35-064805.1\colb 4.070\colc 18.2\cold   24\cole   42\colf   42\colg   72\colh 19.90\coli  ?n\colj ...&...&...&... \eol
\cola BR0401-1711\colb 4.227\colc 16.5\cold   42\cole   53\colf   72\colg   91\colh 19.60\coli   n\colj ...&...&...&... \eol
\cola SDSSpJ040550.26+005931.2\colb 4.050\colc 17.0\cold   19\cole   32\colf   32\colg   42\colh 20.60\coli   n\colj ...&...&...&... \eol
\cola BRJ0529-3552\colb 4.172\colc 16.3\cold ...\cole ...\colf ...\colg ...\colh ...\coli   n\colj ...&...&...&... \eol
\cola PMNJ0741-5305\colb 3.743\colc 16.9\cold   32\cole   53\colf   53\colg   91\colh 19.40\coli   n\colj ...&...&...&... \eol
\cola SDSSpJ084455.08+001848.5\colb 3.690\colc 18.1\cold   32\cole   53\colf   53\colg   91\colh 19.40\coli   n\colj ...&...&...&... \eol
\cola SDSSpJ084811.52-001418.0\colb 4.124\colc 16.9\cold   19\cole   32\colf   33\colg   42\colh 20.30\coli y\colj 19.4&24 &4 &-22.0 \eol
\cola SDSSpJ084811.52-001418.0\colb 4.124\colc 17.4\cold ...\cole ...\colf ...\colg ...\colh ...\coli y?\colj ...&...&...&... \eol
\cola SDSSpJ085430.18+004213.6\colb 4.084\colc 17.6\cold   42\cole   72\colf   72\colg  130\colh 19.00\coli   n\colj ...&...&...&... \eol
\cola SDSSJ094822.97+005554.3\colb 3.878\colc 17.6\cold   24\cole   42\colf   42\colg   53\colh 19.90\coli   n\colj ...&...&...&... \eol
\cola SDSSJ094822.97+005554.3\colb 3.878\colc 18.2\cold ...\cole ...\colf ...\colg ...\colh ...\coli n\colj ...&...&...&... \eol
\cola SDSSJ095755.63-002027.5\colb 3.753\colc 16.7\cold   32\cole   53\colf   53\colg   91\colh 19.40\coli n\colj ...&...&...&... \eol
\cola SDSSJ095755.63-002027.5\colb 3.753\colc 17.5\cold ...\cole ...\colf ...\colg ...\colh ...\coli n\colj ...&...&...&... \eol
\cola SDSSJ100151.58-001627.0\colb 3.674\colc 17.0\cold   53\cole  130\colf   92\colg  140\colh 18.40\coli n\colj ...&...&...&... \eol
\cola SDSSpJ101832.46+001436.4\colb 3.826\colc 17.6\cold   24\cole   42\colf   42\colg   72\colh 19.90\coli n\colj ...&...&...&... \eol
\cola SDSSpJ102043.82+000105.8\colb 4.207\colc 16.9\cold   32\cole   42\colf   53\colg   72\colh 19.60\coli n\colj ...&...&...&... \eol
\cola SDSSJ135134.46-003652.2\colb 4.010\colc 17.2\cold   42\cole   92\colf   72\colg  130\colh 18.70\coli n\colj ...&...&...&... \eol
\cola SDSSJ151618.44-000544.1\colb 3.735\colc 17.8\cold   24\cole   42\colf   42\colg   72\colh 19.90\coli n\colj ...&...&...&... \eol
\cola SDSSpJ152443.19+011358.9\colb 4.095\colc 17.3\cold   53\cole  140\colf   92\colg ...\colh 18.20\coli n\colj ...&...&...&... \eol
\cola SDSSpJ154014.57+001854.7\colb 3.830\colc 18.0\cold   42\cole   72\colf   72\colg  130\colh 19.00\coli y?\colj ...&...&...&... \eol
\cola SDSSpJ161926.87-011825.2\colb 3.840\colc 18.1\cold   32\cole   42\colf   53\colg   72\colh 19.60\coli n\colj ...&...&...&... \eol
\cola SDSSpJ165527.61-000619.2\colb 3.990\colc 16.9\cold   71\cole ...\colf  130\colg ...\colh 18.80\coli n\colj ...&...&...&... \eol
\cola [HB89]2000-330\colb 3.773\colc 15.2\cold ...\cole ...\colf ...\colg ...\colh ...\coli n\colj ...&...&...&... \eol
\cola PC2047+0123\colb 3.799\colc 16.5\cold    9\cole   14\colf   15\colg   19\colh 21.40\coli y\colj 19.7&40 &7 &-22.6 \eol
\cola [VH95]2125-4529\colb 3.763\colc 17.3\cold   71\cole  130\colf  130\colg ...\colh 18.40\coli n\colj ...&...&...&... \eol
\cola [VCV96]Q2133-4625\colb 4.180\colc 18.8\cold   71\cole ...\colf  130\colg ...\colh 18.40\coli n\colj ...&...&...&... \eol
\cola BR2212-1626\colb 3.990\colc 15.6\cold ...\cole ...\colf ...\colg ...\colh ...\coli n\colj ...&...&...&... \eol
\cola BR2212-1626\colb 3.990\colc 16.3\cold ...\cole ...\colf ...\colg ...\colh ...\coli n\colj ...&...&...&... \eol
\cola BRJ2317-4345\colb 3.943\colc 16.6\cold ...\cole ...\colf ...\colg ...\colh ...\coli n\colj ...&...&...&... \eol

\enddata

\tablenotetext{a}{Range of {\it upper limits} for exponential hosts {\it with no evolution}--2mag evolution would lower L's by 6x}
\tablenotetext{b}{Range of {\it upper limits} for deVaucouleurs models {\it with no evolution}--2mag evolution would lower L's by 6x}
\tablenotetext{c}{Observed K magnitude of galaxy corresponding to min $L_{exp}/\lstar$ in column 4}
\tablenotetext{d}{Estimate of host luminosity for disks with and without 2mag of evolution}
\tablenotetext{e}{Host absolute magnitude in the rest frame assuming 2 mag of evolution} 
\end{deluxetable}




\end{document}